\documentclass[final,5p,times,twocolumn,authoryear]{elsarticle}

\usepackage{amssymb}
\usepackage{amsmath}
\usepackage{multirow}
\usepackage[table,xcdraw]{xcolor}
\usepackage{acronym}
\usepackage{float}
\usepackage[table]{xcolor}
\usepackage{makecell}
\usepackage{listings}
\definecolor{delim}{RGB}{20,105,176}
\definecolor{numb}{RGB}{106, 109, 32}
\definecolor{string}{rgb}{0.64,0.08,0.08}
\lstdefinelanguage{json}{
    numbers=left,
    numberstyle=\scriptsize,
    frame=single,
    showspaces=false,
    showstringspaces=false,
    breaklines=true,
    basicstyle=\ttfamily\scriptsize,
    stringstyle=\color{string},
    literate=
     *{0}{{{\color{numb}0}}}{1}
      {1}{{{\color{numb}1}}}{1}
      {2}{{{\color{numb}2}}}{1}
      {3}{{{\color{numb}3}}}{1}
      {\{}{{{\color{delim}{\{}}}}{1}
      {\}}{{{\color{delim}{\}}}}}{1}
      {[}{{{\color{delim}{[}}}}{1}
      {]}{{{\color{delim}{]}}}}{1},
}

\usepackage{microtype}
\usepackage[hidelinks]{hyperref}
\hypersetup{
pdftitle={NASP: Network Slice as a Service Platform for 5G Networks},
pdfauthor={Felipe Hauschild Grings, Gustavo Zanatta Bruno, Lucio Rene Prade, José Marcos Camara Brito, Cristiano Bonato Both}
}

\usepackage{xurl} % Better URL line breaks in bibliography
\usepackage{flushend} % Balance last page columns
\acrodef{5G}{Fifth Generation}
\acrodef{6G}{Sixth Generation}
\acrodef{3GPP}{Third Generation Partnership Project}
\acrodef{AMF}{Access and Mobility Management Function}
\acrodef{CI/CD}{Continuous Integration/Continuous Delivery}
\acrodef{CSMF}{Communication Service Management Function}
\acrodef{eMBB}{Enhanced Mobile Broadband}
\acrodef{IETF}{Internet Engineering Task Force}
\acrodef{IEEE}{Institute of Electrical and Electronics Engineers}
\acrodef{IaC}{Infrastructure as Code}
\acrodef{IoT}{Internet of Things}
\acrodef{mMTC}{Massive Machine-Type Communication}
\acrodef{NFV}{Network Functions Virtualization}
\acrodef{NASP}{Network Slice as a Service Platform} 
\acrodef{NEST}{Network Slice Type}
\acrodef{NG-RAN}{Next Generation Radio Access Network}
\acrodef{NS}{Network Slicing}
\acrodef{NSaaS}{Network Slice as a Service}
\acrodef{NSMF}{Network Slice Management Function}
\acrodef{NSSMF}{Network Slice Subnet Management Function}
\acrodef{NRF}{Network Repository Function}
\acrodef{SBA}{Service-Based Architecture}
\acrodef{SLA}{Service Level Agreement}
\acrodef{SLS}{Service Level Specification}
\acrodef{SDN}{Software-Defined Networking}
\acrodef{SMF}{Session Management Function}
\acrodef{UPF}{User Plane Function}
\acrodef{UDM}{Unified Data Management}
\acrodef{UDP}{User Datagram Protocol}
\acrodef{UE}{User Equipment}
\acrodef{URLLC}{Ultra-Reliable Low Latency Communications}
\acrodef{VPN}{Virtual Private Network}
\acrodef{GSMA}{Global System for Mobile Communications Association}
\acrodef{ZSM}{Zero-touch Network and Service Management}
\acrodef{5GC}{5G Core}
\acrodef{B5G}{Beyond Fifth Generation}
\acrodef{ACM}{Association for Computing Machinery}
\acrodef{ONAP}{Open Network Automation Platform}
\acrodef{OSM}{Open Source MANO}
\acrodef{VNF}{Virtual Network Function}
\acrodef{OpenSlice}{OpenSlice Framework}
\acrodef{AI}{Artificial Intelligence}
\acrodef{E2E}{End-to-End}
\acrodef{DLT}{Distributed Ledger Technologies}
\acrodef{NSI}{Network Slice Instance}
\acrodef{NSSI}{Network Slice Subnet Instance}
\acrodef{ETSI}{European Telecommunications Standards Institute}
\acrodef{O-RAN}{Open Radio Access Network}
\acrodef{RN1}{Radio Access Network Interface}
\acrodef{TN1}{Transport Network Interface}
\acrodef{CN1}{Core Network Interface}
\acrodef{NEF}{Network Exposure Function}
\acrodef{SMO}{Service Management and Orchestration}
\acrodef{Free5gc}{Free 5G Core}
\acrodef{ONOS}{Open Network Operating System}
\acrodef{JSON}{JavaScript Object Notation}
\acrodef{NSST}{Network Slice Subnet Template}
\acrodef{NST}{Network Slice Template}
\acrodef{YAML}{YAML Ain't Markup Language}
\acrodef{HTTP}{Hypertext Transfer Protocol}
\acrodef{VLAN}{Virtual Local Area Network}
\acrodef{REST}{Representational State Transfer}
\acrodef{API}{Application Programming Interface}
\acrodef{IP}{Internet Protocol}
\acrodef{KubeAPI}{Kube-API Server}
\acrodef{Helm}{Helm Chart Package Manager}
\acrodef{Mininet}{Mininet Virtual Network Emulator}
\acrodef{Calico}{Calico}
\acrodef{Multus}{Multus}
\acrodef{non3GPP}{Non-3GPP}
\acrodef{IPsec}{Internet Protocol Security}
\acrodef{OCloud}{O-Cloud Environment}
\acrodef{RAN}{Radio Access Network}
\acrodef{NF}{Network Function}
\acrodef{TN}{Transport Network}
\acrodef{vCPU}{Virtual CPU}
\acrodef{RAM}{Random Access Memory}
\acrodef{CPU}{Computer Processing Unit}
\acrodef{S-NSSAI}{Single Network Slice Selection Assistance Information}
\acrodef{ITU-T}{International Telecommunication Union Telecommunication Standardization Sector}
\acrodef{K8s}{Kubernetes}
\acrodef{CGROUPS}{Control Groups}
\acrodef{PDU}{Protocol Data Unit}
\acrodef{GTP-5G}{GPRS Tunneling Protocol for 5G}
\acrodef{5G-TOURS}{5G-TOURS Project}
\acrodef{5GZORRO}{5GZORRO Project}
\acrodef{5Growth}{5Growth Project}
\acrodef{5G-COMPLETE}{5G-COMPLETE Project}
\acrodef{PCF}{Policy Control Function}
\acrodef{AUSF}{Authentication Server Function}
\acrodef{NSSF}{Network Slice Selection Function}
\acrodef{NWDAF}{Network Data Analytics Function}
\acrodef{QoS}{Quality of Service}
\acrodef{RU}{Radio Unit}
\acrodef{DU}{Distributed Unit}
\acrodef{CU}{Centralized Unit}
\acrodef{CN}{Core Network}
\acrodef{NFVI}{Network Functions Virtualization Infrastructure}
\acrodef{MANO}{Management and Orchestration}
\acrodef{DRL}{Deep Reinforcement Learning}
\acrodef{DCAE}{Data Collection, Analytics, and Events}
\acrodef{CNN}{Convolution Neural Network}
\acrodef{RNN}{Recurrent Neural Network}
\acrodef{DQN}{Deep Q-Network}
\acrodef{CNF}{Cloud-native Network Function}
\acrodef{KPI}{Key Performance Indicator}
\acrodef{RL}{Reinforcement Learning}
\acrodef{GAN}[GAN]{Generative Adversarial Network}
\acrodef{LSTM}[LSTM]{Long Short-term Memory}
\acrodef{SFaaS}{Slice Federation-as-a-Service}
\acrodef{SFC}{Service Function Chain}
\acrodef{NFVO}{Network Function Virtualization Orchestrator}
\acrodef{NSD}{Network Slice Descriptor}
\acrodef{GST}{Generic Network Slice Template}     
\acrodef{TMF}{TeleManagement Forum}
\acrodef{gNB}[gNB]{next-generation Node B}
\acrodef{WAN}[WAN]{Wide Area Network}
\acrodef{UTC}[UTC]{Coordinated Universal Time}
\acrodef{VM}[VM]{Virtual Machine}
\acrodef{NSSAI}[NSSAI]{Network Slice Selection Assistance Information}
\acrodef{SCTP}[SCTP]{Stream Control Transmission Protocol}
\acrodef{NGAP}{NG Application Protocol}
\acrodef{N3IWF}[N3IWF]{Non-3GPP Interworking Function}
\acrodef{SLO}{Service Level Objective}
\acrodef{ML}{Machine Learning}
\acrodef{MOI}{Managed Object Instance}
\acrodef{OSS}{Operations Support System}
\acrodef{BSS}{Business Support System}
\acrodef{NRM}{Network Resource Model}
\acrodef{IOC}{Information Object Class}
\acrodefplural{IOC}{Information Object Classes}
\acrodef{VNFD}{Virtual Network Function Descriptor}
\acrodef{OpenFlow}{OpenFlow Protocol}
\acrodef{sFlow}{sFlow Protocol}
\acrodef{IPFIX}{IP Flow Information Export}
\acrodef{XML}{Extensible Markup Language}
\acrodef{RBAC}{Role-Based Access Control}
\acrodef{UDR}{Unified Data Repository}
\acrodef{RTT}{Round-Trip Time}

\journal{Journal of Network and Computer Applications}

\makeatletter
\def\ps@pprintTitle{%
 \let\@oddhead\@empty
 \let\@evenhead\@empty
 \def\@oddfoot{\footnotesize \copyright~2026 The Authors. Published by Elsevier Ltd. This is an open access article under the CC BY 4.0 license. \hfill}%
 \let\@evenfoot\@oddfoot}
\makeatother

\begin{document}

\begin{frontmatter}

%% Title
\title{NASP: Network Slice as a Service Platform for 5G Networks}

%% Authors with affiliation
\author[inst1]{Felipe Hauschild Grings}
\author[inst2]{Gustavo Zanatta Bruno\corref{cor1}}
\author[inst1]{Lucio Rene Prade}
\author[inst2]{José Marcos Camara Brito}
\author[inst1]{Cristiano Bonato Both}

\cortext[cor1]{Corresponding author. \newline \textit{E-mail address:} gustavo.zanatta@posdoc.inatel.br (G.Z. Bruno). \vspace{1em} \newline \url{https://doi.org/10.1016/j.jnca.2026.104479} \newline Received 18 November 2025; Received in revised form 4 March 2026; Accepted 18 March 2026}

\affiliation[inst1]{organization={Universidade do Vale do Rio dos Sinos - UNISINOS},%
            addressline={Cristo Rei}, 
            city={São Leopoldo},
            postcode={93.022-750}, 
            state={Rio Grande do Sul},
            country={Brazil}}

\affiliation[inst2]{organization={Instituto Nacional de Telecomunicações (INATEL)},%
            addressline={Av. João de Camargo, 510 - Centro}, 
            city={Santa Rita do Sapucaí},
            state={MG},
            postcode={37536-001},
            country={Brazil}}

%% Abstract
\begin{abstract}

With the rapid global adoption of fifth-generation (5G) mobile telecommunications, the demand for highly flexible private networks has surged. A key beyond‑5G feature is network slicing, where the 3rd Generation Partnership Project (3GPP) defines three main use cases: massive Machine-Type Communications (mMTC), enhanced Mobile Broadband (eMBB), and Ultra-Reliable Low-Latency Communications (URLLC), along with their associated management functions. Similarly, the European Telecommunications Standards Institute (ETSI) provides the Zero-Touch Network and Service Management (ZSM) standard, enabling operation without human intervention. However, current technical documents lack definitions for end-to-end (E2E) management and integration across domains and subnet instances. We present a Network Slice as a Service Platform (NASP) that is agnostic to 3GPP and non-3GPP networks, addressing this gap. The NASP architecture comprises (i) onboarding requests for new slices at the business level, translating them into definitions of physical instances and interfaces among domains, (ii) a hierarchical orchestrator coordinating management functions, and (iii) communication interfaces with network controllers. Our NASP prototype is developed based on technical documents from 3GPP and ETSI, analyzing design overlaps and gaps across different perspectives. Results demonstrate the platform's adaptability in handling diverse requests via the Communication Service Management Function. Evaluation indicates that the Core configuration accounts for 68\% of the time required to create a Network Slice Instance. Tests reveal a 93\% reduction in data session establishment time when comparing URLLC and Shared scenarios, i.e., an eMBB-type slice that reuses existing control-plane Network Functions to represent a best-effort provisioning baseline. Finally, we present cost variations for operating the platform with the orchestration of five and ten slices, showing a 112\% variation between Edge and Central deployments.

\end{abstract}

%% Graphical abstract (currently empty; include a file if available)
%\begin{graphicalabstract}
%\includegraphics{grabs}
%\end{graphicalabstract}

% Research highlights
%\begin{highlights}
%\item NASP platform automates end-to-end slice design, deployment, and assurance.
%\item Layered service manager chain links business intents to slice resources.
%\item Prototype merges Kubernetes control, software defined transport, cloud core.
%\item Ultra reliable low latency slice setup cuts session time by 93 percent.
%\item Cost model reveals 112 percent monthly gap between edge and central sites.
%\end{highlights}

%% Keywords
\begin{keyword}
5G \sep Network slicing \sep Network Slice as a Service Platform (NASP) \sep End-to-end (E2E) management \sep Hierarchical orchestrator \sep Edge computing
\end{keyword}

\end{frontmatter}

%% Main text: Include your sections (provided in the first file) 
\section{Introduction}

The rapid, sustained increase in mobile data traffic, combined with the emerging diversity of service requirements from \ac{mMTC} terminals and vertical enterprises, challenges the traditional mobile‐broadband paradigm \citep{Navarro-Ortiz2020}. Legacy networks, initially designed with only a limited set of parameters (e.g., priority class and \ac{QoS} for broadband services), now face the task of accommodating heterogeneous and, in many cases, conflicting operational needs \citep{Vila2020}. For instance, a railway network demands seamless, high-speed mobility management along predefined routes, whereas an electricity metering system requires support for static, low-volume transmissions. Recent advances in \ac{NFV} and \ac{SDN} have paved the way for innovations such as \ac{NS}, which enables the logical separation of network functions and resources tailored to specific technical and commercial requirements \citep{Wijethilaka2021,Ksentini2020,Mahdi2023}.

\ac{NS} has conceptual roots in earlier technologies, such as the 802.1Q virtual local‑area network standard~\citep{IEEE8021Q-2018} and \ac{IETF} RFC 4026 \acp{VPN} \citep{rfc4026}, which provided isolated broadcast and session domains. However, deploying \ac{NS} in cellular environments introduces additional challenges, including mobility management, control‑plane authentication, and session and charging management in the user plane \citep{mahyoub2024security}. Studies led by the \ac{3GPP} and initiatives by major industry players indicate an evolution toward \ac{E2E} network slices that integrate both \ac{3GPP} and non‑\ac{3GPP} domains, enhancing connectivity for the \ac{IoT} and enabling diverse service customization \citep{Nguyen2016, bertenyi2018ng, Linnartz2022-lb}. Despite these advancements, the complexity stemming from distributed interfaces across \ac{RAN}, \ac{TN}, and \ac{CN} necessitates novel orchestration and management solutions \citep{Wyszkowski2024, 7962822}.

The evolution toward heterogeneous mobile networks, driven by massive \ac{IoT} deployments, emerging business-oriented services, and stringent performance requirements, motivates the need for a comprehensive, automated approach to \ac{NS}. Existing studies have addressed resource allocation using machine‑learning techniques \citep{Jiang, Li2018DeepRL} and have developed frameworks for \ac{NS} control and assurance \citep{Abbas2020IBNSlicingIN, Theodorou, Bega}. However, fully integrated \ac{E2E} orchestration that translates \ac{SLA}/\ac{SLS} requirements into practical \ac{NS} designs remains an open challenge, as clear guidelines that bridge the gap between business-level intents and concrete slice configurations spanning the \ac{RAN}, \ac{TN}, and \ac{CN} domains are still lacking in the literature.

The central research question addressed in this work is how standardized components can be orchestrated and integrated to provide \ac{NSaaS}. We argue that effective orchestration and integration must be achieved through well‑defined interfaces among diverse network components and a clear hierarchy of responsibilities within the orchestrator. In this context, we propose \ac{NASP}, which (i) translates business‑rule templates into subnet‑instance descriptors, (ii) applies hierarchical slice‑deployment strategies, and (iii) integrates closed‑loop management functions by leveraging \ac{CI/CD} and \ac{IaC} practices. \ac{NASP} manages \ac{E2E} slices across both \ac{3GPP} and non‑\ac{3GPP} networks seamlessly.

% Contributions paragraph  
The main contributions of this work are threefold. First, we introduce \ac{NASP}, an automated \ac{NSaaS} platform that translates business‐rule templates into multi‐domain slice descriptors and orchestrates \ac{E2E} deployments across both \ac{3GPP} and non‐\ac{3GPP} networks. Second, we propose a hierarchical orchestration model, including \ac{CSMF}, \ac{NSMF}, and domain‐specific \acp{NSSMF}, which integrate closed‐loop quality assurance via \ac{CI/CD} and \ac{IaC} practices, enabling dynamic \ac{SLA} compliance through AI‐driven telemetry analytics. Third, we develop and release a fully functional prototype built on Kubernetes, \ac{ONOS}, my5G-RANTester~\citep{Silveira2022}, and \ac{Free5gc}, demonstrating seamless integration, extensibility, and native support for physical and virtual resources in heterogeneous deployment environments.

% Impactful results paragraph  
Our comprehensive evaluation, covering \ac{mMTC}, \ac{URLLC}, shared, and non-\ac{3GPP} use cases, yields several impactful findings. 
In this context, the shared use case corresponds to an \ac{eMBB}-type slice that reuses existing control-plane \acp{NF}, serving as a best-effort baseline for comparison against fully isolated deployments.
We observe that 66\% of the \ac{E2E} slice instantiation time is spent on Core domain configuration, pinpointing critical optimization targets. \ac{URLLC} slices achieve up to a 93\% reduction in data-session establishment time compared to shared slices. A prototype closed-loop quality assurance module demonstrates timely detection and mitigation of performance anomalies in our testbed. Finally, a cost-efficiency analysis reveals a 112\% variation in monthly operational expenses between edge and centralized deployments, highlighting the economic differences between the two deployment options.

The remainder of this article is organized as follows. Section~\ref{sec:background} provides essential background on \ac{NFV}, \ac{SDN}, and \ac{NS} technologies. Section~\ref{sec:rw} reviews related work and highlights existing gaps in E2E \ac{NS} orchestration. Section~\ref{sec:proposal} details the proposed \ac{NASP} architecture and its key design choices. Section~\ref{sec:prototype} describes the implementation and prototype development of the platform. Section~\ref{sec:evaluation} outlines the methodology used to assess platform performance, while Section~\ref{sec:results} presents the experimental results and analysis. Finally, Section~\ref{sec:conclusion} concludes the article and discusses directions for future research.

\section{Background and Evolution of 5G Networks}\label{sec:background}

\begin{figure}[!h]
  \centering
  \includegraphics[width=\linewidth]{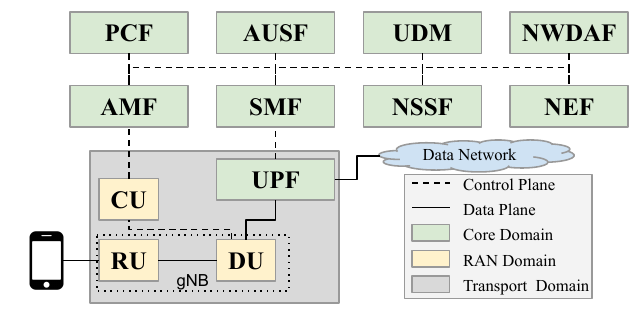}
  \caption{\ac{5G} Core architecture.}
  \label{fig:5g_architecture}
\end{figure}

The evolution of mobile telecommunications toward \ac{5G} introduced network programmability and \ac{NSaaS} to better meet emerging service and market needs \citep{3gpp.21.915, afolabi2018network}. According to \citet{3gpp.23.501}, the \ac{5G} system and its \ac{SBA} were first specified in Releases~15--16, with continued refinements in later \ac{3GPP} releases (TS~23.501 v19.2.0, Rel-19). The architecture retains core elements, \ac{UE}, \ac{NG-RAN}, and \ac{5GC}, and separates control and user planes via \ac{AMF} and \ac{UPF}. Figure~\ref{fig:5g_architecture} highlights supporting functions such as \ac{PCF} (policy), \ac{AUSF} (authentication), \ac{NSSF} (slice selection), \ac{NWDAF} (analytics), and \ac{NEF} (exposure).

\ac{NS} is a central \ac{5G} innovation, allowing multiple logical networks to share a common infrastructure while meeting distinct \acp{SLA}. Each slice orchestrates resources across the \ac{RAN}, \ac{TN}, and \ac{CN} and is characterized according to \ac{3GPP} and \ac{GSMA} guidance \citep{3gpp.23.501, gsma-116} (GSMA~NG.116 v10.0). The \ac{NEST} concept groups key attributes, throughput, latency, and reliability, into a service profile. For example, an \ac{eMBB} slice may target 99.999\% availability, multimedia telephony, specific \ac{QoS}, and strong session continuity. Moreover, isolation is essential since congestion or scaling in one slice must not degrade others, and separation must hold for performance, management, and security/privacy \citep{3gpp.28.552}. In this case, the system can be viewed in three domains, as can be seen in Figure~\ref{fig:5g_architecture}: the RAN (gNB with \ac{RU}/\ac{DU}/\ac{CU}), the \ac{TN} (access--core transport), and the \ac{CN} (interfacing with external Data Networks via the \ac{UPF}).

\begin{figure}[!h]
  \centering
  \includegraphics[width=\linewidth]{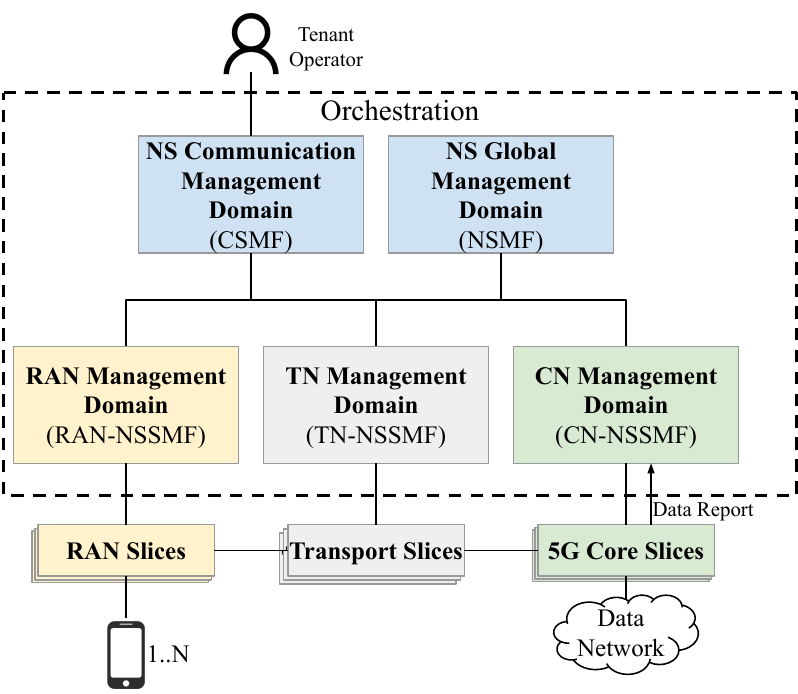}
  \caption{\ac{NS} orchestration architecture.}
  \label{fig:ns-architecture}
\end{figure}

Releases~17--18 \citep{3gpp.21.917,3gpp.21.918} extend slicing toward \ac{NSaaS}, enabling operators to offer tenants customized, on-demand slices. \ac{NSaaS} exposes management capabilities via standardized \acp{API} from \ac{3GPP} and TM~Forum frameworks \citep{gsma-116}. Slice management relies on \ac{CSMF}, \ac{NSMF}, and \ac{NSSMF} to translate tenant demands into an \ac{E2E} deployment. Figure~\ref{fig:ns-architecture} outlines this orchestration, which includes  an \ac{NS} communication interface and a management domain comprising the \ac{NSMF}, \ac{CSMF}, and domain-specific entities (RAN-\ac{NSSMF}, TN-\ac{NSSMF}, 5GC-\ac{NSSMF}). These coordinate instantiation, monitoring, and resource allocation across the \ac{NG-RAN}, \ac{TN}, and \ac{5GC}. Multiple slices (1--N) can coexist with dedicated resources and connect to external Data Networks through defined channels.

\begin{figure*}[!h]
  \centering
  \includegraphics[width=.95\linewidth]{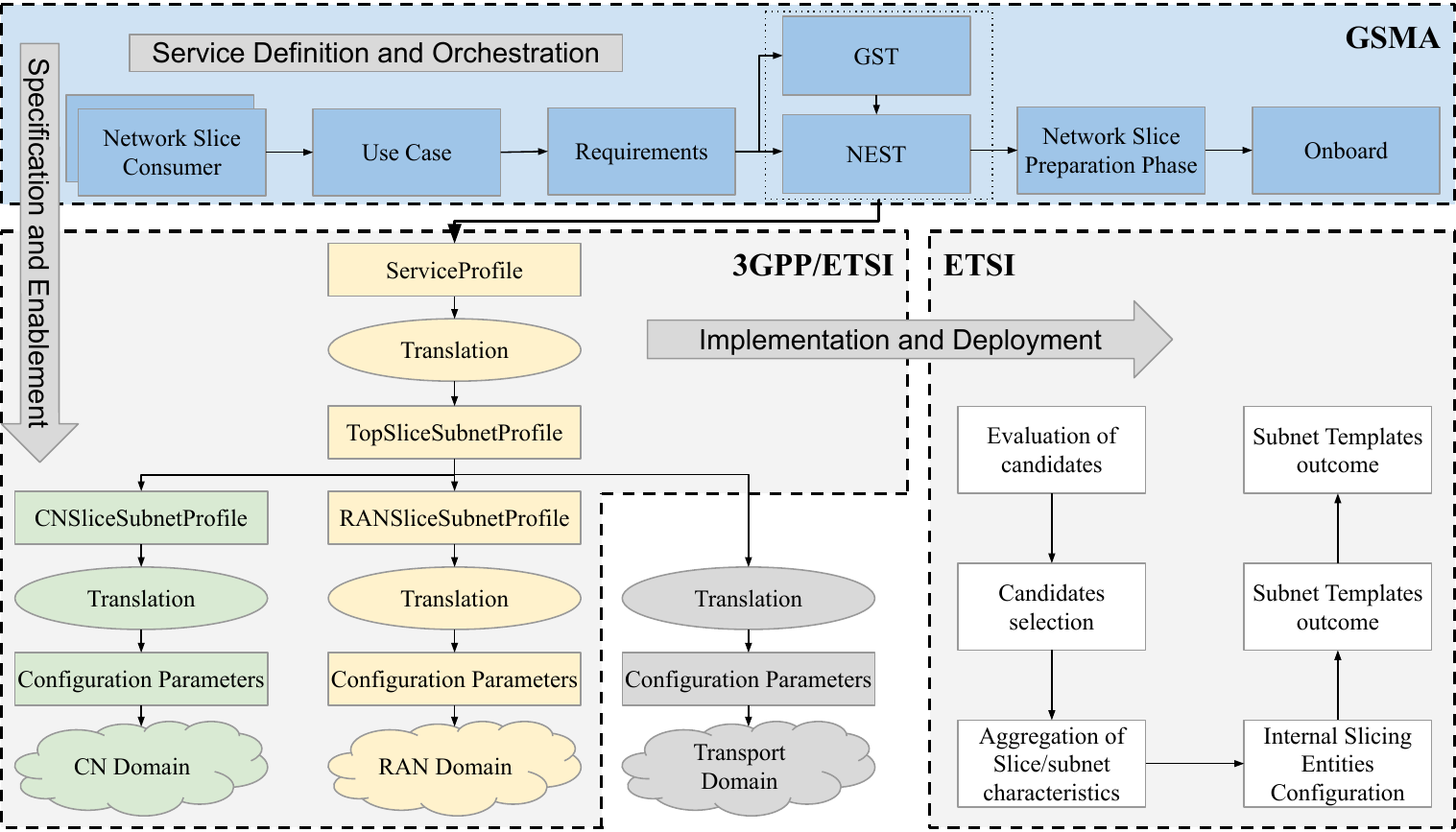}
  \caption{End-to-End \ac{NS} Management and Orchestration Across GSMA, 3GPP/ETSI, and ETSI Standards. Automation Scopes (AS1--AS4) follow the ETSI~ZSM model for the evaluation, selection, configuration, and deployment of slice subnets.}
  \label{fig:standarts-overview}
\end{figure*}

\ac{NS} lifecycle management, from preparation (blueprints, reservations) to operation (activation, monitoring, capacity planning) and decommissioning, benefits from modularity and isolation \citep{5gcorebook-ns}. The resulting operational complexity motivates \ac{ZSM}, which applies automation, \ac{ML}, and \ac{AI} for self-configuration, self-monitoring, self-healing, and self-optimization \citep{etsi-zsm}. Therefore, beyond-\ac{5G} systems build on these principles to deliver agile, customized services that meet diverse \acp{SLA}.
Building on these advances, the latest \ac{3GPP} specifications (Rel-18/19) outline a precise \ac{NRM} utilizing \acp{IOC} (e.g., \texttt{NetworkSlice}, \texttt{NetworkSliceSubnet}) to standardize management \citep{3gpp.21.918}.
These are augmented with \texttt{ServiceProfile} and \texttt{SliceProfile} descriptors that align with \ac{GSMA} \ac{GST} to maintain globally consistent \ac{SLA} representations.
By codifying \ac{NRM} objects, states, and operations across multi-vendor definitions (\ac{XML}, \ac{JSON}, YANG), these standards provide the necessary foundation for \ac{E2E} slice orchestration and exposure of \ac{NSaaS} through open \acp{API}.
The instantiation of a \ac{NSI} follows a standardized \ac{API} sequence among \ac{OSS}/\ac{BSS}, \ac{CSMF}, \ac{NSMF}, domain \acp{NSSMF}, \ac{NFV}--\ac{MANO}, and the Inventory. Figure~\ref{fig:standarts-overview} illustrates the high-level flow and the interplay between \ac{GSMA}, \ac{3GPP}/\ac{ETSI}, and \ac{ETSI} standards in this process. The process begins when the \ac{OSS}/\ac{BSS} captures a tenant request encoded with \ac{GSMA} \ac{GST}/NG.116, translating commercial requirements into a service order as specified by \citet{GSMA-NG116}. The \ac{CSMF} produces a \texttt{ServiceProfile} that maps high-level \ac{SLA} attributes (throughput, latency, coverage, isolation) into a standards-compliant representation \citep{3gpp.28.541}.

Following \ac{GST} modeling, the \ac{NSMF} translates the \texttt{ServiceProfile} into an \texttt{allocateNsi} request to decompose the slice into domain-specific subnets.
Each domain \ac{NSSMF} resolves the \texttt{SliceProfile} and engages the \ac{NFV}--\ac{MANO} stack to instantiate resources, which are persisted as \acp{MOI} via \texttt{createMOI} \cite{3gpp.28.622, ETSI-NFV-MANO}.
When dynamic adjustments are necessary, runtime parameter tuning is achieved through \texttt{modifyMOIAttributes} \citep{3gpp.28.623}.
This structured progression from business intent (\ac{GST} $\rightarrow$ \texttt{ServiceProfile} $\rightarrow$ \texttt{SliceProfile}) to runtime \acp{MOI} corresponds directly to the \ac{ETSI} \ac{ZSM} Automation Scopes AS1--AS4 \citep{etsi-zsm}, enabling a repeatable, zero-touch orchestration process.

\section{Related Work}\label{sec:rw}

\begin{table*}[!ht]
\centering
\caption{\small Representative Related Work.}
\label{tab:rw-table}
\resizebox{\linewidth}{!}{%
\begin{tabular}{l|c|c|c|c|c|c|c}
\hline
\rowcolor[HTML]{EFEFEF}
\textbf{Works} 
& \textbf{Type} 
& \makecell{\textbf{B5G/AI}\\\textbf{Ready}}
& \makecell{\textbf{Project}\\ \textbf{Model}}
& \textbf{Architecture}
& \makecell{\textbf{Service}\\ \textbf{Automation}}
& \makecell{\textbf{\ac{E2E}}\\ \textbf{Service}\\ \textbf{LC Mgmt.}}
& \textbf{NSaaS} \\
\hline
\citet{ONAP}            & Project  & L & H & H & M & M & L \\ \hline
\citet{OSM}              & Project  & L & H & H & M & M & L \\ \hline
\citet{5GTours}      & Project  & L & H & H & M & M & L \\ \hline
\citet{5g-zorro}      & Project  & M & M & M & L & L & L \\ \hline
\citet{Baranda2020}   & Project  & L & H & H & L & L & L \\ \hline
\citet{tranoris2021openslice} & Project  & M & H & H & M & M & L \\ \hline
% ---------- ⬇ NEW OPEN-SOURCE PROJECTS ⬇ ----------
\citet{camara2025}        & Project  & M & H & – & H & – & M \\ \hline
\citet{opencapif2025}     & Project  & M & H & M & M & L & L \\ \hline
\citet{nephio2025}     & Project  & H & H & H & H & M & M \\ \hline
\citet{sylva2024}        & Project  & M & H & H & M & L & L \\ \hline
\citet{hexaX2} & Project & H & M & H & H & M & M \\ \hline
\citet{monb5g} & Project & H & M & M & H & M & L \\ \hline
\citet{daemon} & Project & H & M & H & H & M & L \\ \hline
% ---------- ⬆ NEW OPEN-SOURCE PROJECTS ⬆ ----------
\citet{Li2018DeepRL}                 & Article  & L & H & L & L & L & L \\ \hline
\citet{Jiang}                     & Article  & M & H & L & H & M & L \\ \hline
\citet{Bega}                       & Article  & M & H & L & L & H & L \\ \hline
\citet{Abbas2020IBNSlicingIN}     & Article  & M & H & M & H & M & L \\ \hline
\citet{Fernandez}             & Article  & M & H & M & M & L & L \\ \hline
\citet{Theodorou}             & Article  & H & H & L & H & H & L \\ \hline
\citet{Larrea2023}               & Article  & M & H & L & L & M & L \\ \hline
\citet{Scotece2023-5gkube}      & Article  & M & M & M & H & L & L \\ \hline
\citet{mlopez2023mapek} & Article & H & L & M & H & M & L \\ \hline
\citet{Wyszkowski2024}       & Article  & M & L & H & L & L & L \\ \hline
\citet{Dalgitsis2024-kp}      & Article  & H & H & M & M & H & L \\ \hline
\citet{Esmat2024}                 & Article  & H & H & H & M & H & L \\ \hline
\citet{Chowdhury2024}         & Article  & H & H & H & H & M & L \\ \hline
\citet{mlops2024} & Article & H & L & H & H & H & M \\ \hline
\citet{zhao2025}                   & Article  & H & H & H & H & M & L \\ \hline
\rowcolor[HTML]{EFEFEF}\textbf{NASP (This Work)} & Article & M & H & H & H & H & H \\ \hline
\end{tabular}%
}
\\[0.4em]
\small Maturity: Low (L), Medium (M), High (H); “–” = not applicable.
\end{table*}

The recent literature on \ac{NS} for \ac{5G} and beyond networks has centered on both \ac{E2E} slice design and dynamic orchestration. Table~\ref{tab:rw-table} provides a qualitative comparison of the surveyed solutions. The 'Works' column lists each solution, and \textit{Type} distinguishes between runnable open-source projects and concept-focused articles. The \textit{B5G/AI Ready} score reflects how well a study incorporates forward-looking \ac{B5G} concepts and \ac{AI}-driven automation, from legacy \ac{5G} (L) to full \ac{AI}/\ac{ML} integration (H). \textit{Project Model} rates the maturity of released software artifacts, ranging from a fully maintained public repository (H) through partial or prototype code (M) to no public code (L). \textit{Architecture} captures the breadth of the \ac{E2E} slice design across \ac{RAN}, \ac{TN}, and \ac{CN} domains. \textit{Service Automation} records adherence to \ac{ETSI} \ac{ZSM} closed-loop orchestration principles, and \textit{\acs{E2E} Service LC Mgmt.} measures the degree of automation across preparation, instantiation, assurance, and decommissioning. Finally, \textit{\ac{NSaaS}} indicates whether the work exposes \ac{NS} as a tenant-facing service, progressing from no exposure (L), to domain-specific offerings (M), and, at the top end, full multi-domain \ac{NSaaS} capability (H).

Several open-source initiatives automate \ac{5G} slice orchestration but vary in scope. \citet{ONAP} and \citet{OSM} offer comprehensive \ac{MANO} frameworks but require significant integration effort for full \ac{E2E} automation. Building on these, subsequent projects such as \citet{5GTours}, \citet{5g-zorro}, and \citet{Baranda2020} demonstrated vertical use cases but reported limited architectural evolution.
Since 2023, the landscape has shifted toward \ac{B5G} and \ac{AI}-native solutions. Large-scale research programmes such as \citet{hexaX2} investigate \ac{B5G} architectures that emphasize \ac{AI}-driven automation and sustainability. Projects such as \citet{monb5g} and \citet{daemon} develop \ac{AI}/\ac{ML}-powered monitoring and control loops, aligning with the \ac{ETSI} \ac{ZSM} vision. The MAPE-K framework introduced by \citet{mlopez2023mapek} is now a widely cited pattern for self-adaptive systems. Moreover, the MLOps practices cataloged by \citet{mlops2024} are being integrated into network automation to streamline the lifecycle of \ac{AI} models in production networks. Furthermore, community-driven initiatives such as \citet{camara2025}, \citet{opencapif2025}, \citet{nephio2025}, and \citet{sylva2024} develop open, cloud-native \ac{API} and automation frameworks backed by major operators and vendors.

Pushing lifecycle automation further, OpenSlice, as described by \citet{tranoris2021openslice}, adopts TM Forum’s Open Digital Architecture to drive catalog-based order decomposition that supports zero-touch onboarding, scaling, and retirement of \ac{NFV} artifacts. However, it delegates granular \ac{5GC} tasks, such as \ac{UPF} placement, \ac{PCF}/\ac{NRF} affinity, and network-exposure \acp{API}, to external \ac{MANO} stacks, excluding the \ac{SBA} from the reconfiguration loop. In response, the \ac{NASP} framework presented in this article unifies hierarchical slice composition, \ac{AI}-powered assurance, and tenant-facing \ac{NSaaS} exposure across \ac{RAN}, \ac{TN}, and \ac{CN} domains.

\ac{AI}-driven research is steadily closing specific gaps in the \ac{NS} lifecycle. \citet{Li2018DeepRL} were the first to apply \ac{DRL} for joint radio- and core-resource balancing under fluctuating loads. \citet{Jiang} introduced intelligence slicing, a unified framework that instantiates \ac{AI} modules, such as \ac{RNN}-based channel prediction and security-anomaly detection, on demand. \citet{Bega} integrated a 3-D \ac{CNN} capacity-forecasting model (DeepCog) with \ac{RL} control loops, enabling proactive \ac{VNF} scaling throughout the slice lifecycle. \citet{Abbas2020IBNSlicingIN} proposed IBNSlicing, an intent-based, \ac{E2E} framework that automatically translates high-level slice “contracts” into \ac{OSM} and FlexRAN templates; a hybrid \ac{GAN}, i.e.,  \ac{LSTM} + \ac{CNN}, forecasts slice-level CPU and RAM requirements to trigger proactive scaling and admission decisions. Moreover, \citet{Fernandez} designed a \ac{DLT}-enabled \ac{5G} marketplace in which multiple providers securely advertise, negotiate, and trade heterogeneous resources via smart-contract automation, validating the concept with a Corda-based \ac{VNF}-trading proof of concept. Complementing these point solutions, \citet{Wyszkowski2024} proposed a taxonomy of network-slice-instance patterns that now guides subsequent work.

\begin{figure*}[!ht]
    \centering
    \includegraphics[width=\textwidth]{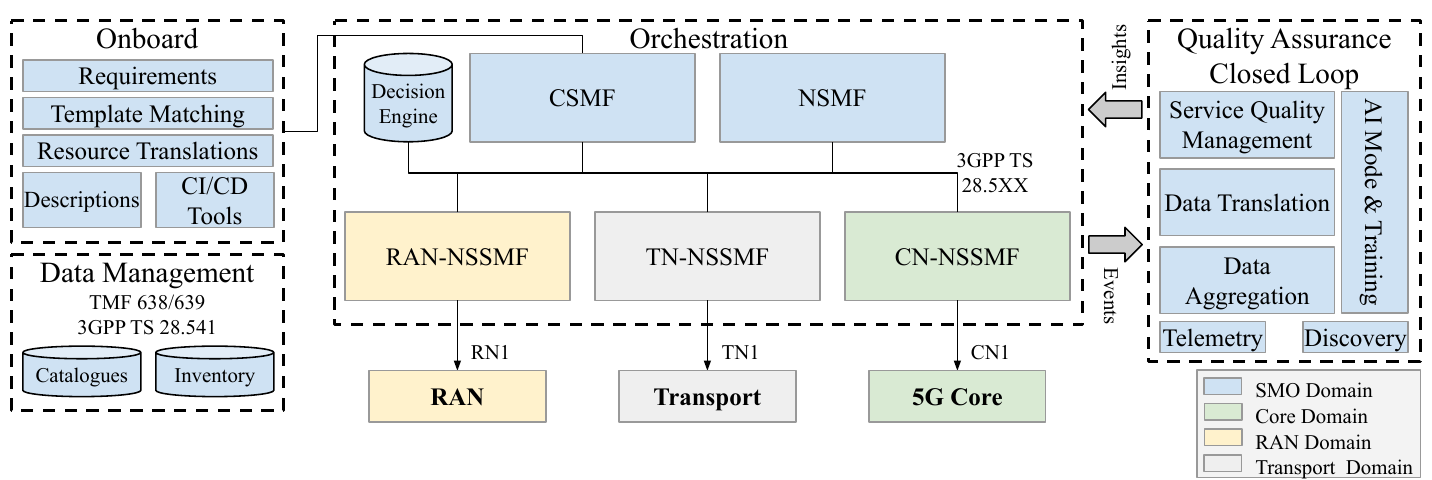}
    \caption{\ac{NASP} architecture.}
    \label{fig:architecture}
\end{figure*}

Extending the state of the art, \citet{Theodorou} outlined a blockchain-based, zero-touch service-assurance loop for cross-domain slicing, in which smart-contract-bound \acp{SLA} feed \ac{AI}-driven monitoring, prediction, and automated reconfiguration across the lifecycle. \citet{Dalgitsis2024-kp} proposed a cloud-native \ac{SFaaS} framework that augments the \ac{GSMA} Operator Platform with a new east–west \acp{API}, enabling seamless slice federation across operators and demonstrating the approach on a live \ac{5G}/\ac{B5G} testbed. \citet{Esmat2024} decomposed carrier-grade \acp{SLA} into federated slice objectives, whereas \citet{Chowdhury2024} introduced Accelerator, an intent-driven, zero-touch resource-slicing algorithm that places and schedules \ac{SFC} workloads for \ac{B5G} and legacy services across \ac{SDN}/\ac{NFV} infrastructures.

Cloud-native platforms are likewise paving the way for zero-touch cores. \citet{Scotece2023-5gkube} presented 5G-Kube, a Kubernetes-native \ac{CNF} stack that cuts slice instantiation times below 200 ms. \citet{Larrea2023} developed CoreKube, a cloud-native mobile core whose stateless, message-focused workers are containerized and orchestrated by Kubernetes. Additionally, \citet{zhao2025} combined adaptive \ac{RAN} slicing with hierarchical \ac{DRL}, achieving \ac{E2E} \acsp{KPI} across \ac{RAN}, \ac{TN}, and \ac{CN}.
Moreover, \citet{Wyszkowski2024} present a detailed tutorial that addresses a blind spot in slicing research and standardization: the design-time organization of slice and subnet instances. They systematize a taxonomy of slicing entities and formalize the slice/subnet design process, clarifying concepts not uniformly covered in 3GPP or ETSI. Building on this, the authors enumerate the essential information items a complete design must capture and propose a standards-aligned structure to record them. The tutorial decomposes design into fine-grained, ordered activities and offers automation guidelines that evolve from assisted handling to fully unattended, declarative workflows capable of dynamic slicing. 

Despite these state of the art advances, three challenges remain: (i) \ac{E2E} designs that span \ac{RAN}, \ac{TN}, and \ac{CN} within a \ac{SBA}; (ii) fully closed-loop slice lifecycle management aligned with \ac{SLA} objectives; and (iii) an operational \ac{NSaaS} framework that unifies these capabilities. The \ac{NASP} framework addresses these gaps by (i) translating \ac{GSMA} business templates into domain-specific descriptors, (ii) orchestrating hierarchical slice instantiation across \ac{RAN}, \ac{TN}, and \ac{CN} via open interfaces, such as \ac{RN1}, \ac{TN1}, and \ac{CN1}; and (iii) embedding \ac{AI}-driven closed-loop assurance. Compared with other solutions, \ac{NASP} integrates native \ac{B5G} support, run-time optimization, and a demonstrable \ac{NSaaS} prototype, illustrating a viable path toward zero-touch, \ac{ZSM}-compliant \ac{B5G} deployments. It is worth noting that while the \ac{NASP} framework architecture is designed to integrate advanced \ac{AI}-driven closed-loop assurance, we rate its current \textit{B5G/AI Ready} status as Medium (M) (see Table~\ref{tab:rw-table}) to reflect that the current prototype relies mainly on rule-based policies, with full \ac{AI} integration planned for future work.
\section{NSaaS Platform Overview}\label{sec:proposal}

%% coloquei a figura an seção anterior para ficar no top da págica que começa essa seção.

This section introduces \ac{NASP}, which aligns with concepts from \ac{3GPP}, \ac{ETSI}, and \ac{GSMA}frameworks to enable \ac{E2E} slicing. The platform mediates between business intent and network implementation by automatically translating high‑level business templates into \ac{NS} definitions, enabling multi‑domain resource orchestration and monitoring.
In this context, \ac{NASP} translates \ac{GSMA} business templates into versioned \acp{NSST} through a hierarchical deployment model.
Once initially processed by the Onboard module, business requests are handled by the \ac{CSMF}, where they are transformed into service requirements and forwarded to the \ac{NSMF}, which assigns slice identifiers (IDs) and delegates domain‑specific tasks.
Standardized interfaces (\ac{RN1}, \ac{TN1}, and \ac{CN1}) ensure seamless integration across the \ac{RAN}, \ac{TN}, and \ac{CN}.
Automation is achieved using \ac{CI/CD} and \ac{IaC} practices, while a closed‑loop monitoring system enforces continuous telemetry‑based \ac{SLA} compliance.

Figure~\ref{fig:architecture} presents the \ac{NASP} reference architecture, comprising five main elements. The \textbf{\textit{Onboard module}} (i) translates tenant requests into domain-specific requirements, leveraging \ac{GSMA} and \ac{3GPP} standards, as well as operator-specific parameters, to ensure that high-level business templates are accurately mapped to the corresponding slice (\ac{NST}/\acp{NSST}) definitions. The \textbf{\textit{Data Management module}} (ii) stores and manages all slice descriptors and artifacts through a domain-agnostic catalog and inventory, ensuring version control, lifecycle tracking, and consistent deployment across network domains. The \textbf{\textit{Orchestration module}} (iii), fulfilling the roles of both \ac{CSMF} and \ac{NSMF}, coordinates slice instantiation across multi‑domain networks and incorporates functions such as data translation and resource mapping, as depicted by the integrated Resource Translations block. The \textbf{\textit{Quality Assurance Closed Loop module}} (iv) monitors performance via closed‑loop telemetry and employs \ac{AI}‑driven analytics to detect anomalies, interfacing with \ac{CI/CD} tools to trigger automated reconfigurations that maintain \ac{SLA} compliance. Finally, the \textbf{\textit{Interfaces}} (v) \ac{RN1}, \ac{TN1}, and \ac{CN1} ensure connectivity among domain controllers, facilitating uniform management and synchronization across physical and virtual infrastructures.

\subsection*{\textbf{Onboard module}}

This \ac{NASP} module acts as the single northbound entry point through which tenants submit their slice requests encoded using the \ac{GSMA} \ac{GST}, thereby managing the \textit{design-time} (or preparation) phase.
Listing~\ref{lst:slice_request} presents a minimal excerpt of a customized \ac{eMBB} slice request, demonstrating how tenant requirements, such as \ac{SLA} constraints and domain-specific topological limits, are encoded and submitted.

\begin{lstlisting}[language=json, caption={Example of a custom tenant slice request encoded as a JSON object.}, label={lst:slice_request}, float=h]
{ "name": "Custom 5G Network Slice",
  "NST": {
    "type": "custom",
    "Slice Attributes": {
      "availability": 1,
      "Supported Data Network": "internet",
      "SSQ": {
        "Packet Delay Budget": 0.00012,
        "Packet Error Rate": 0.0000001,
        "Maximum Data Burts Volume": 0.001},
      "UE density": 10000},
    "resource_description": {
      "core": { "nfs": [{"name": "amf"}, {"name": "smf"}, {"name": "upf"}] },
      "ran": { "nfs": [{"name": "ueransim", "type": "gnb", "replicas": 2}] },
      "tn": { "routes": [{"name": "backhaul"}] }
    }}}
\end{lstlisting}

Incoming service intents first pass through the Requirements micro‑service, where \ac{SLA} parameters such as latency, throughput, and availability are normalized.
The data stream is forwarded to Template Matching, whose rule‑based engine correlates the normalized attributes with a \ac{3GPP}‑compliant library of versioned \acp{NSST}, selecting the candidate that minimizes resource footprint while maximizing reuse.
If no existing template mapping can satisfy the normalized attributes, the Onboard module rejects the request and returns an error to the tenant.
Once the descriptors are prepared and validated, the Onboard module triggers the \textit{run-time} instantiation phase by forwarding the processed request to the Orchestration module (specifically linking to the \ac{CSMF}).

Once a suitable template is chosen, Resource Translations converts logical requirements into concrete descriptors for virtual compute, storage, radio, and transport resources, using mapping rules derived from \cite{TMF638,TMF639} and \cite{3gpp.28.541}. The translated artifact is forwarded to the Descriptions service, which renders a standards‑compliant \ac{NSD} in \ac{YAML} format, directly consumable by the orchestration tier. A GitOps‑driven \ac{CI/CD} pipeline performs syntax validation, policy checks, and version tagging before committing the artifact to an immutable registry.
During onboarding, each \ac{GST} attribute, such as \texttt{e2eLatency}, \texttt{serviceAvailability}, is persisted as a \texttt{ServiceSpecification/SlaTarget} object via \ac{TMF}~638 (v5.0.0).
These targets are version‑controlled alongside the \ac{NS} descriptors and immediately promoted to \ac{TMF}~640 \ac{SLO} records so that downstream orchestration and assurance functions can reference a single, immutable source of truth.

\subsection*{\textbf{Data Management module}}

\begin{figure*}[!h]
    \centering
    \includegraphics[width=\linewidth]{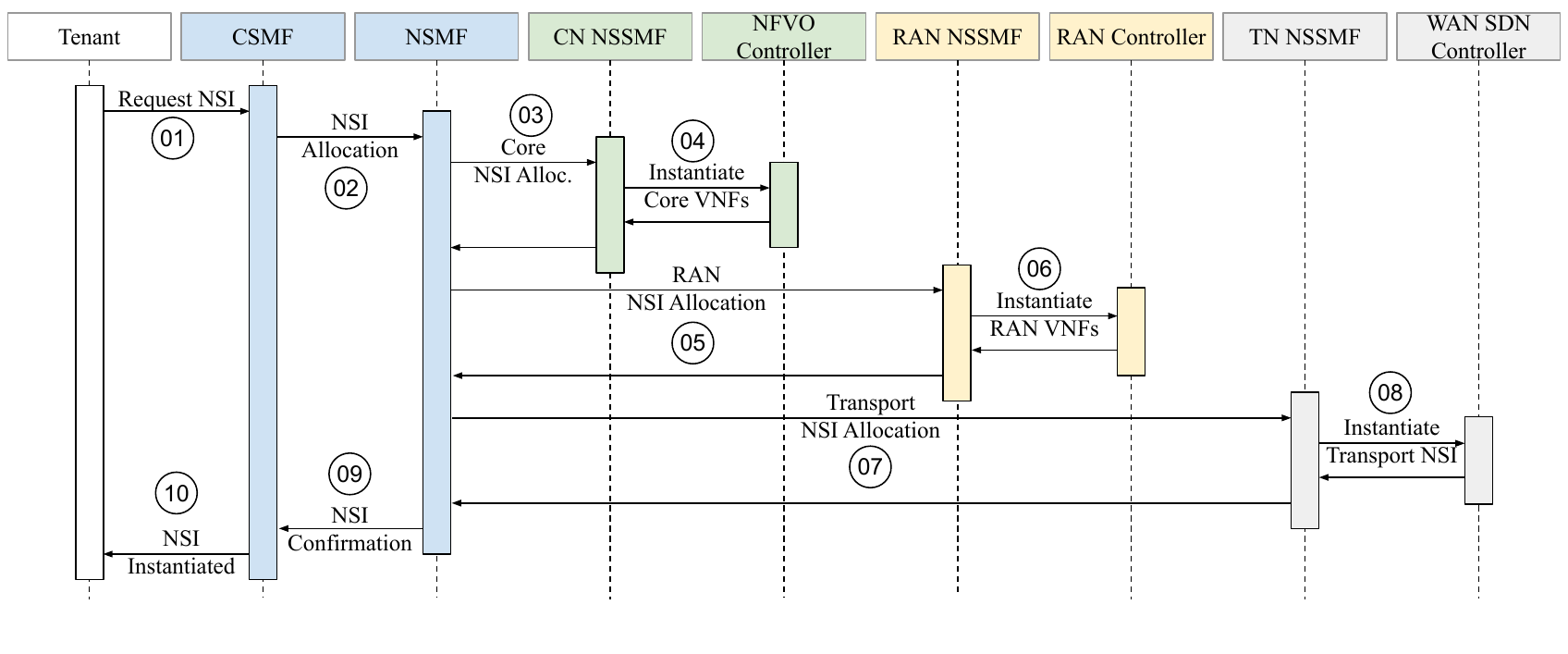}
    \caption{Sequence diagram of slice resource allocation.}
    \label{fig:sequence-diagram}
\end{figure*}

All slice descriptors and artifacts are stored in a domain-agnostic Catalog, which maintains versioned \acp{NSD}, \acp{NSST}, virtual network function images, and \ac{Helm} charts. This repository provides template definitions for domain-specific orchestrators, ensuring that instantiation across domains follows consistent and deterministic patterns. A complementary Inventory does not duplicate these descriptors. Instead, it tracks the current lifecycle state of each slice instance across the underlying physical and virtual resources. It exposes authorised state information to admission control and assurance functions. Leveraging these resources, the Onboard module translates the tenant's business intent into service definitions that can be processed by the orchestration tier without human intervention. The Catalog serves as the design-time repository for immutable templates, such as \acp{NSD}, \acp{NSST}, \acp{VNFD}, and \ac{Helm} charts, validated and version-controlled through \ac{CI/CD}. The Inventory captures the runtime state of instantiated \acp{NSI} and \acp{NSSI}, including lifecycle attributes, \acp{KPI}, and \ac{SLA} compliance. Overall, the Catalog defines what can be deployed, and the Inventory records what is deployed. This separation preserves traceability without duplication and supports federation through abstracted, authorized views.

\subsection*{\textbf{Orchestration module}}

A four-tier hierarchy outlines responsibilities: Tier-1 (\ac{CSMF}) captures business \ac{SLA}s and converts them into service requirements; Tier-2 (\ac{NSMF}) assigns slice identifiers and orchestrates tasks across domains; Tier-3 domain-specific \acp{NSSMF}, \ac{RAN}, \ac{TN}, and \ac{CN}, allocate resources and instantiate slices from the descriptors stored in the Catalog; and Tier-4 comprises the actual physical or virtual instances configured from those descriptors. Rather than duplicating data-management functions, the orchestrator treats the Catalog solely as a source of template information and queries the Inventory for the live state of slice resources during provisioning and monitoring. 

The end-to-end \textit{run-time} workflow, illustrated in Figure~\ref{fig:sequence-diagram}, begins when the tenant's slice request is forwarded by the Onboard module to the \ac{CSMF}~(01); the request is then forwarded to the \ac{NSMF}, which reserves an \ac{S-NSSAI}, retrieves the corresponding slice templates and engages the \ac{CN}~\ac{NSSMF} to deploy the required \ac{CN} \acp{VNF} under the supervision of the \ac{NFVO} Controller~(02–04).
Once an \ac{AMF} endpoint is available, the \ac{RAN}~\ac{NSSMF} instantiates the \ac{gNB} components through the \ac{RAN} Controller so that they attach to the correct \ac{AMF}~(05–07).
After both \ac{RAN} and \ac{CN} domains are operational, the \ac{TN}~\ac{NSSMF} instructs the \ac{WAN} \ac{SDN} Controller to establish transport connectivity, reports completion to the \ac{NSMF} and triggers the final activation confirmation that is relayed to the tenant via the \ac{CSMF}~(08–10).
All configuration artifacts are stored in the Inventory through idempotent transactions, guaranteeing that repeated executions converge to a consistent slice state, and the final slice configuration is preserved for subsequent monitoring and lifecycle management.

% 
% Removi pq achei redundante com o parágrafo abaixo, que é mais completo.

%Upon receipt of the slice request, the \ac{CSMF} converts the \ac{TMF}~638 (v5.0.0) \texttt{SlaTargets} into \ac{TMF}~640 (v5.0.0) \ac{SLO} instances and publishes them on the Quality‑Assurance message bus. The \ac{NSMF} translates each \ac{SLO} into concrete \ac{KPI} thresholds and injects them into the closed‑loop policy engine. Domain-specific \acp{NSSMF} embeds these thresholds within their local controllers, allowing breaches to be detected at the lowest possible latency. Violations generate \ac{TMF}~642 (v5.0.0) events, which in turn trigger automated scaling, relocation, or healing workflows through \ac{TMF}~645 (v5.0.0) Assurance APIs.

% removi pq isso não é um módulo da arquitetura
%\subsection*{\textbf{Business‑SLA Derivation \& Enforcement}}

The transformation from high‑level business commitments into enforceable artifacts occurs in four logical phases leveraging \ac{TMF} Open APIs (v5.0.0).
(i) \textit{Production:} \ac{GST} attributes are normalized and stored as \ac{TMF}~638 \texttt{SlaTarget} objects, each of which is linked, via a mapping table inside the Resource Translations micro‑service, to the corresponding \ac{TMF}~639 resource identifiers, e.g., transport bandwidth pools or radio‑spectrum slices.
(ii) \textit{Decomposition:} the \ac{CSMF} converts these targets into \ac{TMF}~640 \ac{SLO} records and distributes them to the \ac{NSMF} and the domain‑specific \acp{NSSMF}, which embed the derived \ac{KPI} thresholds in their local controllers.
(iii) \textit{Enforcement:} during operation, the Quality‑Assurance loop continuously compares real‑time telemetry with the injected thresholds; any breach raises a \ac{TMF}~642 notification that automatically triggers scaling, relocation, or healing workflows through \ac{TMF}~645 Assurance APIs.
(iv) \textit{Tenant visibility:} current \ac{SLA} status is exposed through a \ac{TMF}~645 /assurance/v1/monitorings endpoint, and tenants may subscribe to \ac{NEF} webhooks for proactive breach alerts.

The orchestration workflow follows the 3GPP management and service specifications, including TS 28.531 and TS 28.541 for slice lifecycle coordination, as well as TS 23.501 and TS 38.413 for inter-domain reference points. While 3GPP does not mandate a domain instantiation order, the \ac{NASP} prototype deploys \ac{CN} → \ac{RAN} → \ac{TN} to mirror runtime dependencies.
\ac{CN} is instantiated first, so the \ac{AMF} can expose its control- and user-plane endpoints: N2, i.e., \ac{NGAP} over \ac{SCTP}, for signalling and registration, and N3, i.e., \ac{GTP-5G} over \ac{UDP} for user-plane traffic. The \ac{AMF} publishes these endpoints via its management \ac{API}, and they are persisted in the Inventory with \ac{IP} addresses and ports.
Next, the \ac{RAN} orchestrator configures \ac{gNB} components to complete NG Setup and register with the selected \ac{AMF} over N2. Once \ac{CN} and \ac{RAN} are active, their artifacts, such as \ac{gNB} \acp{IP}, \ac{VLAN} IDs, and N3 forwarding parameters, are provided to the \ac{TN} orchestrator, which programs slice-aware connectivity via the \ac{SDN} controller, e.g., \ac{OpenFlow} intents and \ac{VLAN}-based paths, for both N2 and N3 traffic between \ac{RAN} and \ac{CN} subnets.
This sequence establishes control-plane signalling before data-plane activation, aligns with the layering hierarchy, and remains compliant with the 3GPP slice-management framework and information models defined in TS 28.541.

\subsection*{\textbf{Quality Assurance Closed Loop module}}

This module supervises every active \ac{NS} through a non‑real‑time control loop that begins with Telemetry Discovery, which retrieves multilayer observability data (counters, alarms, traces, and logs) via the \ac{NEF}, \ac{RAN} monitoring interfaces, and \ac{SDN} probes. Although transport-domain \ac{SDN} telemetry (\ac{OpenFlow} and \ac{sFlow}/\ac{IPFIX}) is not natively covered by \ac{3GPP} TS 28 5xx, a lightweight adaptor inside the Data Translation micro-service enriches every record with the \ac{S-NSSAI} and converts raw counters, such as port-tx\_bytes, queue delay, packet-drop events, into their \ac{3GPP}-compliant counterparts, defined in \citet{3gpp.28.552}. The normalized \acp{KPI} are pushed (i) towards the \ac{NWDAF} for analytics or (ii) as \ac{TMF} 642 events for fault management. This translation keeps the entire assurance loop \ac{3GPP}-compliant while still exploiting the high-resolution visibility provided by \ac{SDN} probes. The records are normalized and time‑stamped before publication on a common bus. Data Aggregation aligns the heterogeneous streams on the slice ID, such as S‑NSSAI, and enriches them with business \acp{KPI} from the catalog. Finally, Data Translation harmonizes vendor‑specific semantics, converting resource counters into utilization metrics, mobility events into experience indicators, and sharing information into isolation scores, while mapping slice intents onto the configuration vocabulary of the \ac{SMO}, \ac{NSMF}, and domain‑specific \acp{NSSMF}.

Service Quality Management can host a hierarchy of supervised and deep-\ac{RL} models that at the decision tier can forecast \ac{QoS} degradation, detect anomalies, and compute corrective actions such as scaling, function relocation, or policy updates, all guarded by \ac{SLA}-aware safety checks. \ac{AI} Mode \& Training aims to orchestrate the learning lifecycle, retraining models on curated snapshots when concept drift appears and promoting validated artifacts to the online inference path without service disruption. Outcomes are disseminated through Events and Feeds, which stream real-time notifications to the operations dashboard, publish periodic assurance reports to tenants, and immutably log every action with before-and-after metrics to close the loop and enable continual improvement.

\subsection*{\textbf{Interfaces}}

The \ac{NASP} control plane exposes three domain‑agnostic southbound interfaces that enable uniform \ac{NS} lifecycle automation across the \ac{RAN}, \ac{TN}, and \ac{CN} domains. Each interface conveys an intent encoded as a \ac{YAML} document that augments the \ac{3GPP} network‑resource model in TS 28.541 (v19.4.0, Rel‑19) with \ac{GST} attributes defined by \ac{GSMA} NG.116 (v10.0). \ac{RN1} targets the radio‑access network, \ac{TN1} the transport network, and \ac{CN1} the mobile core.
For example, \ac{RN1} submits a \ac{RAN} Slice Set to the \ac{K8s} \ac{API} server managing the cloud‑native \ac{RAN}. The manifest is deployed via \ac{Helm} releases, Network Attachment Definitions, and accompanying custom resources that configure the required \ac{gNB} functions. Moreover, \ac{TN1} delivers a Path Intent over a RESTful \ac{HTTP}/2 channel to the \ac{SDN} controller. Because this intent is technology-agnostic, the controller can translate it into the appropriate southbound protocols based on the underlying network equipment. For instance, it may employ \ac{OpenFlow}~1.5 flow‑mod messages for fine-grained traffic engineering on virtual switches, or push YANG-modeled configurations via Netconf to manage physical transport routers. Finally, \ac{CN1} forwards a \ac{CN} Slice Set to the \ac{K8s} \ac{API} server that orchestrates the \ac{CN}. Therefore, the manifest is applied as \ac{Helm} releases and ancillary resources, including a Data Plane Policy object that configures the user‑plane function datapath.

\section{Prototype Implementation}\label{sec:prototype}

This section presents the proof-of-concept built to validate the proposed \ac{NASP} architecture for beyond-\ac{5G} mobile networks. The prototype implements a context‑aware slice orchestrator in a containerized environment i.e., \ac{NASP}’s Onboard, Orchestration, and Quality Assurance Closed Loop modules manage the end-to-end slice lifecycle (creation, update, and termination). It further extends the \ac{SMO} concept by embedding business‑level objectives from \ac{GSMA} templates directly into the orchestration workflow, bridging business intent and network service delivery.
The prototype runs on Kubernetes v1.28 (Ubuntu 20.04 LTS) to manage containerized network functions across the \ac{RAN}, \ac{TN}, and \ac{CN} domains. my5G-RANTester modules provide \ac{RAN} functions, \ac{ONOS} supervises a \ac{Mininet} virtual environment to emulate the \ac{TN}, and \ac{Free5gc} implements the \ac{CN}. The \ac{NASP} prototype source code is available at: \url{https://github.com/fhgrings/NASP}.

%The overall architecture is depicted in Fig.~\ref{fig:architecture}.

The infrastructure uses Docker and the Kubernetes \ac{API} server’s \ac{REST} interface for configuration and management. The \ac{NASP} control plane stores \ac{JSON}-formatted \acp{NSST} and \acp{NST} in a document‑oriented database. Infrastructure descriptors are supplied through \ac{Helm} charts and \ac{YAML} manifests. The orchestrator applies context-aware logic that maps tenant metadata (e.g., latency budgets, UE density) to infrastructure-level configurations, dynamically adapting Helm chart parameters during instantiation. Monitoring, tracing, and alerting are provided by Prometheus, Istio, and Grafana, respectively. Control modules are written in Python 3.9; Flask handles \ac{HTTP} requests, and Django serves a single‑page Web interface. Bash scripts coordinate low‑level Linux operations across domains.

\begin{figure}[!h]
    \centering
    \includegraphics[width=\linewidth]{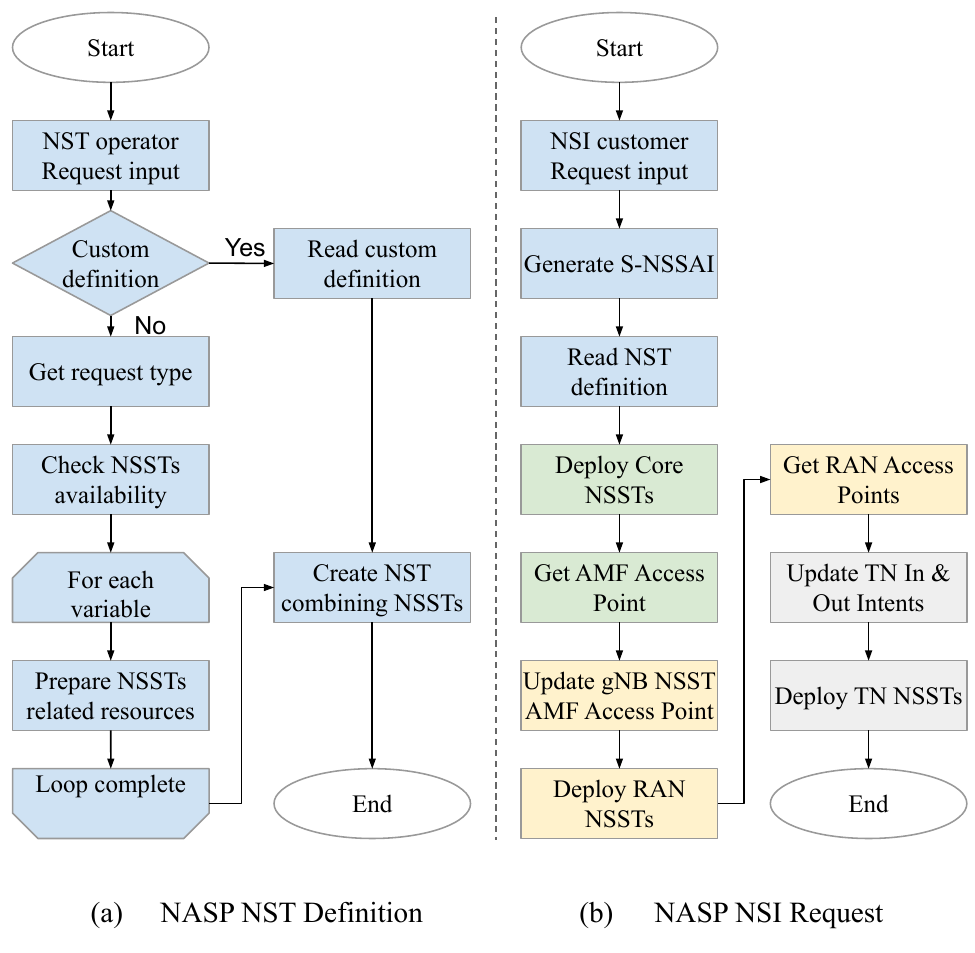}
    \caption{NASP workflows: (a) \ac{NST} definition and (b) \ac{NSI} instantiation.}
    \label{fig:nst-nsi-fluxograms}
\end{figure}

Figure~\ref{fig:nst-nsi-fluxograms}(a) depicts the \ac{NASP} \ac{NST} definition workflow, triggered when an operator submits a slice-template request. The orchestrator first checks for a tenant-provided definition; if present, the \ac{YAML}/\ac{Helm} fragment is parsed and validated. Otherwise, the engine derives the slice type (\ac{eMBB}, \ac{URLLC}, \ac{mMTC}, or an operator-specific profile) from request metadata. The catalog is queried for versioned \acp{NSST} per domain (\ac{RAN}, \ac{TN}, \ac{CN}) together with version tags and resource footprints. For each exposed variable (e.g., maximum \ac{UE} density, latency budget, sharing policy), the engine instantiates parameter sets and reserves artifacts (container images, \ac{Helm} values, ConfigMaps). Finally, the selected \acp{NSST} are composed into an \ac{E2E} \ac{NST} by generating a parent chart that imports the sub-charts and exports a consolidated values file. The resulting artifact is stored and marked as \textit{Ready} for subsequent instantiation.

Figure~\ref{fig:nst-nsi-fluxograms}(b) outlines the \ac{NSI} instantiation workflow. When a customer requests an operational slice, the orchestrator generates a unique \ac{S-NSSAI}, associates it with the tenant ID, and retrieves the corresponding \ac{NST} produced in phase (a). Deployment proceeds domain-wise. First, \ac{CN} \acp{NSST} are instantiated to initialize control-plane functions and obtain the \ac{AMF} access-point address. This address is then injected into the \ac{RAN} slice template so that the \ac{gNB} can bootstrap with the correct \ac{AMF} target. Next, the orchestrator queries the inventory for \ac{RAN} anchor points, updates ingress and egress intents on the \ac{TN} controller, and deploys the \ac{TN} \ac{NSST}, which installs slice-specific forwarding rules. The workflow completes once both \ac{RAN} and \ac{TN} \acp{NSST} report readiness, enabling \acp{UE} to attach and exchange data through the customized transport paths.

\begin{figure}[!h]
    \centering
    \includegraphics[width=\linewidth]{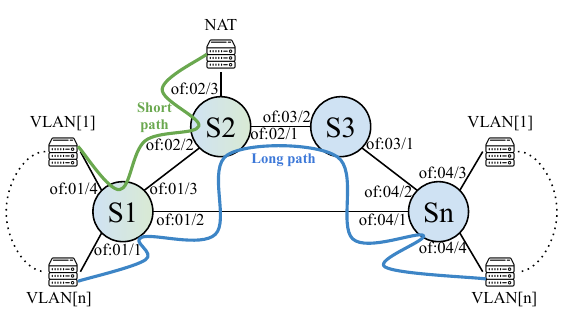}
    \caption{Emulated \ac{TN} topology with \ac{VLAN}-based segmentation.}
    \label{fig:transport-topology}
\end{figure}

\ac{NASP} manipulates \ac{IP} tables inside Kubernetes pods for \ac{TN} configuration.
Each pod’s network interfaces are dynamically reconfigured to route traffic using \ac{VLAN}-based segmentation, providing isolation and meeting performance requirements (Figure~\ref{fig:transport-topology}).
Although the \ac{TN} is programmed via \ac{ONOS}/OpenFlow, every packet retains a lightweight 802.1Q tag.
Distinct \ac{VLAN} IDs encode slice-specific \ac{QoS} policies: a short path prioritizes low latency for \ac{URLLC}, whereas a longer path provides redundancy for \ac{mMTC} and Shared slices.
These tags act as in-band slice and \ac{QoS} markers, allowing switches to differentiate flows locally, reduce flow-table pressure, and perform deterministic path selection before any controller interaction.
This mechanism preserves layer-2 isolation and remains compatible with legacy Ethernet segments~\citep{IEEE8021Q-2018}.

\subsection*{\textbf{Implemented NASP Modules and Scope}}

\textit{Onboard module:} Supports (i) intent ingestion via \ac{REST}/\ac{JSON}, (ii) slice-profile derivation from \ac{GSMA}-aligned service templates (\ac{eMBB}, \ac{URLLC}, \ac{mMTC}, non-\ac{3GPP}), (iii) catalog lookup of versioned \acp{NSST}, (iv) variable expansion and Helm-based template synthesis, and (v) persistence of rendered \acp{NST} in a document-oriented store.

\textit{Orchestration module:} Implemented with a hierarchical chain consistent with \ac{CSMF}$\rightarrow$\ac{NSMF}$\rightarrow$\ac{NSSMF}. Each domain \ac{NSSMF} (\ac{RAN}, \ac{TN}, and \ac{CN}) exposes idempotent deploy/update/delete operations. Cross‑domain sequencing (\ac{CN}$\rightarrow$\ac{RAN}$\rightarrow$\ac{TN}) and dependency injection, such as passing the \ac{AMF} endpoint to the \ac{gNB} chart, are operational.

\textit{Quality Assurance Closed Loop module:} Implements telemetry collection (Prometheus exporters), \ac{KPI} extraction (deployment time, attach latency, availability, resource consumption), and threshold- or rule-based corrective actions such as \ac{AMF} reconfiguration. Although the architecture includes components for \ac{AI}-driven analytics, the current implementation relies primarily on rule-based mechanisms, leaving advanced \ac{AI} capabilities (e.g., predictive scaling, anomaly classification) for future work.

\textit{Interfaces:} Implemented (\ac{RN1}, \ac{TN1}, \ac{CN1}) as a unified northbound \ac{REST} layer that normalizes slice intents into domain‑specific manifests (\ac{Helm}/\ac{YAML} for Kubernetes, \ac{JSON} payloads for \ac{ONOS}). Adaptors translate abstract bandwidth/latency/shareability fields into concrete \ac{CPU} quotas, pod replica counts, and flow‑rule priorities.

We consider Security and Policy Hooks as a basic enforcement mechanism via Kubernetes \ac{RBAC} and namespace scoping, plus slice-specific \ac{VLAN} tagging and (for non-\ac {3GPP}) \ac{IPsec} tunnel establishment through the \ac{N3IWF}. Fine‑grained policy engines (\ac{AI}‑driven admission) are future work.

\subsection*{\textbf{Service Scenarios and Network-Function Mapping}}

We instantiated four slice scenarios to produce the results discussed later: (i) \ac{URLLC}, (ii) \ac{mMTC} (a.k.a., mIoT in later figures), (iii) Shared (an \ac{eMBB}-type slice that reuses existing control‑plane \acp{NF} to measure best‑case provisioning), and (iv) non‑\ac{3GPP} access (offload via \ac{N3IWF}). Table~\ref{tab:scenario-mapping} summarizes the architectural deltas that drive the different deployment and performance outcomes.

\begin{table*}[!t]
\centering
\caption{Scenario‑specific mapping of \ac{5G} \acp{NF}, placement, and isolation mechanisms.}
\label{tab:scenario-mapping}
\resizebox{.8\linewidth}{!}{%
\begin{tabular}{l|c|c|c|c}
\hline
\rowcolor[HTML]{EFEFEF}
\textbf{Scenario} & \shortstack{\textbf{\ac{CN}}\\\textbf{Functions}} & \shortstack{\textbf{\ac{RAN}}\\\textbf{Config}} & \shortstack{\textbf{\ac{TN} Path /}\\\textbf{\ac{VLAN}}} & \shortstack{\textbf{Isolation}\\\textbf{Highlights}} \\
\hline
\ac{URLLC} & \shortstack{Dedicated \ac{AMF}, \acs{SMF},\\\ac{UPF} (edge); shared \ac{AUSF},\\ \acs{UDR}, \ac{NRF}, \ac{PCF}} & \shortstack{\ac{gNB} \ac{CU}/\ac{DU}\\at edge; low PHY\\timer values} & \shortstack{Short 4-hop path;\\\ac{VLAN} 101\\(high priority)} & \shortstack{Dedicated control +\\data plane; \ac{CPU} limits;\\exclusive flow rules} \\ \hline
\ac{mMTC} & \shortstack{Shared \ac{AMF}/\acs{SMF};\\lightweight dedicated \ac{UPF};\\enhanced UDR scaling} & \shortstack{Std. \ac{gNB}; extended\\RA backoff (contention\\absorption)} & \shortstack{Long path (+2 hops\\resiliency); \ac{VLAN} 102} & \shortstack{Separate namespace;\\per-pod quotas;\\distinct \ac{S-NSSAI}} \\ \hline
\shortstack{Shared\\(\ac{eMBB})} & \shortstack{Reuses \ac{AMF}/\acs{SMF}/\ac{UPF};\\no new \ac{CN} pods} & \shortstack{Std. \ac{gNB}; default\\scheduler} & \shortstack{Same as \ac{mMTC} path\\(reused flows); \ac{VLAN} 102} & \shortstack{Logical slice via\\\ac{S-NSSAI}; policy \ac{QoS}\\classes} \\ \hline
non-\ac{3GPP} & \shortstack{Adds \ac{N3IWF} + \ac{IPsec};\\dedicated \ac{UPF};\\shared \ac{AMF}/\acs{SMF}} & \shortstack{Minimal \ac{gNB} role\\(Wi-Fi offload focus)} & \shortstack{Short path + tunnel\\endpoint; \ac{VLAN} 104} & \shortstack{\ac{IPsec} tunnel;\\namespace + \ac{VLAN} tag\\+ \ac{S-NSSAI}} \\ \hline
\end{tabular}%
}
%\\[0.4em]
%\small \acf{AMF}; \acf{SMF}; \acf{UPF}; \acf{AUSF}; \acf{UDR}; \acf{NRF}; \acf{PCF}; \acf{N3IWF}; \acf{gNB}; acf{VLAN} Random Access (RA).
\end{table*}

%Provisioning Time (Figure~\ref{fig:deployment-time}): The Shared slice is fastest because no fresh Core \acp{NF} are instantiated; only differential policy objects and flow rules are applied. \ac{URLLC} takes longer due to dedicated \ac{AMF}/\ac{SMF}/\ac{UPF} pods with stricter readiness gates. Non‑\ac{3GPP} adds the \ac{N3IWF} and \ac{IPsec} association, while \ac{mMTC} deploys (or scales) database‑heavy components (UDR interactions), yielding intermediate delays.

%Attach / Registration Latency (Fig.~\ref{fig:ue-conn-time}): \ac{URLLC} benefits from edge placement (intra‑slice round‑trip $<3$ ms across the short \ac{TN} path) and an exclusive \ac{UPF}. Non‑\ac{3GPP} incurs extra control‑plane exchanges for tunnel setup. \ac{mMTC} shows a long tail due to random‑access congestion with many simulated devices. Shared reuses multipurpose \ac{CN} pods whose queueing increases median latency.

%Resource Usage and Cost (Figs.~\ref{fig:slice-usage}, \ref{fig:slice-price}): Dedicated \ac{URLLC} control‑plane replicas raise early \ac{CPU} allocation; Shared amortizes cost by multiplexing; \ac{mMTC} scaling of state backends raises \ac{RAM}; non‑\ac{3GPP} adds moderate overhead for encryption endpoints.

\subsection*{\textbf{Isolation Strategy}}

Isolation across the four scenarios is enforced in complementary layers. (1) \textit{Control‑plane logical isolation:} Each slice obtains a unique \ac{S-NSSAI}. Dedicated slices (\ac{URLLC}) run separate \ac{AMF}/\ac{SMF}/\ac{UPF} pods in isolated Kubernetes namespaces, and shared slices rely on policy separation inside common pods. (2) \textit{Resource isolation:} Kubernetes \ac{CPU}/memory requests and limits reserve deterministic compute shares for dedicated \acp{NF} and \texttt{HorizontalPodAutoscaler} triggers (80\% threshold) let \ac{URLLC} replicas scale independently. (3) \textit{Data‑plane segmentation:} \ac{VLAN} tags (Table~\ref{tab:scenario-mapping}) plus \ac{ONOS}-installed OpenFlow rules steer per‑slice traffic over distinct priority queues and (for \ac{URLLC}) the minimal‑hop path. (4) \textit{Security/access isolation:} non‑\ac{3GPP} traffic traverses an \ac{IPsec} tunnel terminated at the \ac{N3IWF} and slice‑specific Kubernetes \texttt{NetworkPolicies} restrict pod reachability. (5) \textit{State isolation:} Although \ac{AMF}/\ac{SMF} are stateless, per‑slice subscriber and session state is keyed by \ac{S-NSSAI} in the \ac{UDR} and cached via slice‑scoped \texttt{ConfigMaps}, preventing cross‑slice leakage. These details explain the performance differences observed later and provide the architectural context for the four evaluated service scenarios.

\section{Evaluation Methodology}\label{sec:evaluation}

This section describes the methodology used to assess the performance of the proposed \ac{NASP} architecture. The evaluation focuses on statistically robust and cost-aware metrics across five aspects: \ac{E2E} slice-instantiation time, scalability, flexibility, cost efficiency, and \ac{UE} connection latency. We perform the experiments on a cloud-native infrastructure that emulates realistic distributed topologies using public cloud resources, \ac{SDN} \ac{WAN} emulators, and containerized network function deployment.

\subsection*{\textbf{Testbed Infrastructure and Network Topology}}

Experiments were conducted on \acp{VM} provided by DigitalOcean.
The cloud testbed comprises heterogeneous \acp{VM} running Ubuntu~22.04 LTS with the Linux kernel~5.15.0‑88‑generic.
Control‑plane components ran on \textit{s‑4vcpu‑8gb} instances (4 \ac{vCPU}, 8 GB RAM), whereas data‑plane elements used \textit{s‑8vcpu‑16gb} instances (8 \ac{vCPU}, 16 GB RAM).
We dedicated a separate \acp{VM} to \ac{ONAP}, \ac{Mininet} with \ac{ONOS}, and a Kubernetes cluster.
We used Kubernetes~v1.28.3 with \ac{Calico} and \ac{Multus} to support multi‑tenancy, with autoscaling thresholds of 80\% (scale‑up) and 20\% (scale‑down).
Persistent monitoring data were exported to a Prometheus–Grafana stack, and all timestamps were converted to \ac{UTC} to avoid clock‑skew artifacts.

\begin{figure}[!h]
    \centering
    \includegraphics[width=0.8\linewidth]{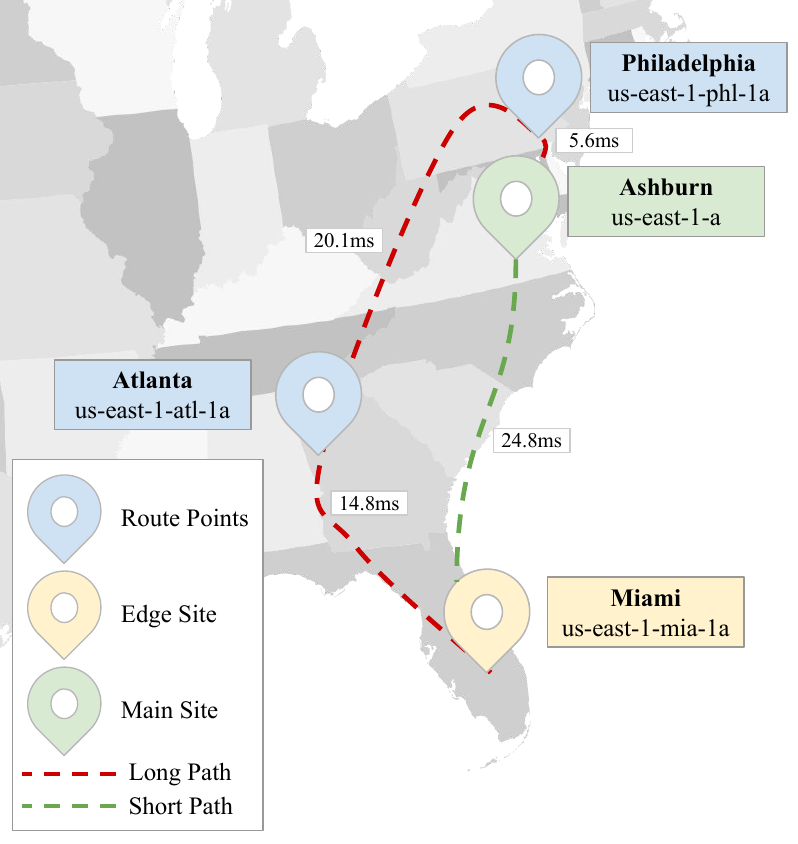}
    \caption{Long‑range cloud‑network topology.}
    \label{fig:topology-mininet}
\end{figure}

In this work, we propose three types of cloud sites for the network performance evaluation: Regional (Central), Metropolitan (Edge), and Internal (Extreme‑Low‑Latency Edge), as defined by \citet{gustavo2024evaluating}. The Internal site offered limited instance types, so we normalized cost comparisons using pricing from all sites. We emulated the network topology using \ac{Mininet} and the \ac{ONOS} controller. Figure~\ref{fig:topology-mininet} shows the long‑range cloud‑network topology, where latency samples were collected from geographically separated regions. Metrics were sampled at 1~Hz over a 72-hour period. A cubic non-linear regression (adjusted \(R^2_{\text{adj}} = 0.93\)) was used to summarize average-latency trends over a representative 24-hour cycle.

\subsection*{\textbf{Performance Metrics}}

The evaluation considers the following five aspects:
\ac{E2E} slice‑instantiation time (i): measured by decomposing design, configuration, and activation phases.  
Scalability (ii): assessed by increasing the number of slices and user connections without observable performance degradation (up to 250 concurrent \acp{UE} per slice).  
Flexibility (iii): inferred from the ability to customize slice requests and operate diverse \ac{5G} network functions profiles, including \ac{eMBB}, \ac{URLLC}, \ac{mMTC}, and non‑\ac{3GPP}.  
Cost efficiency (iv): quantified as the ratio between performance and operational expenses across heterogeneous instance types; we used the AWS TCO calculator to cross-check DigitalOcean pricing.  
\ac{UE} connection latency (v): measured to capture the impact of network management and routing efficiency on end‑user experience. All latency statistics are reported as the median \(\pm\) interquartile range (IQR) to mitigate the influence of outliers.

\section{Experimental Evaluation} \label{sec:results}

This section presents a comprehensive experimental evaluation of the proposed \ac{NASP} solution. The results are organized along two dimensions: (i) the design and deployment of \ac{E2E} \acp{NSI} and (ii) their performance and lifecycle management. In addition to the within-testbed measurements reported in this section, we include a structured cross-study quantitative contextualization. This analysis connects the Related Work discussion with the \ac{NASP} experimental results. The cited solutions differ substantially in architectural scope (e.g., \ac{CN}-focused versus \ac{E2E} \ac{RAN}/\ac{TN}/\ac{CN} orchestration), lifecycle coverage, and experimental setup (e.g., simulation, emulation, real, or hybrid testbeds). Therefore, we discuss representative reported metrics from prior work and preserve their original definitions and units. Furthermore, we qualify each comparison according to its degree of alignment with \ac{NASP} (Direct, Partial, Contextual, or Not comparable). This approach provides quantitative grounding without forcing equivalence when metric semantics, control boundaries, or measurement scope differ.

% Scenario architecture details appear in Table~\ref{tab:scenario-mapping}.
\begin{figure}[h!]
    \centering
    \includegraphics[width=\linewidth]{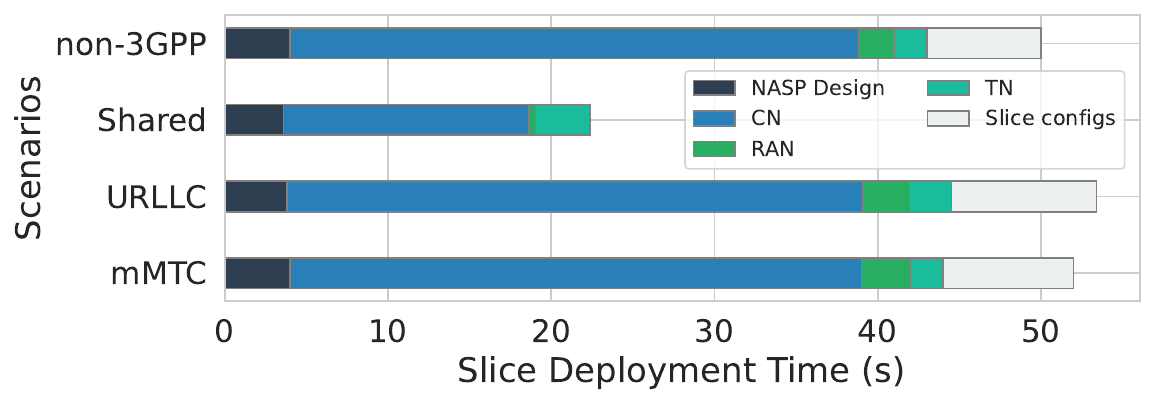}
    \caption{Slice deployment time across four scenarios (URLLC, mMTC, Shared eMBB reuse, non-3GPP offload).}
    \label{fig:deployment-time}
\end{figure}

\subsection{Slice Provisioning and Instantiation Delay}
The \ac{NASP} design-and-deployment evaluation quantifies the time required for slice provisioning. Figure~\ref{fig:deployment-time} shows the breakdown from the initial slice request to full deployment across the \ac{RAN}, \ac{TN}, and \ac{CN} domains, including the subsequent auto-configuration phase. We compare four scenarios (Table~\ref{tab:scenario-mapping}): \ac{URLLC}, \ac{mMTC}, Shared (eMBB reuse), and non-3GPP offload. The Shared slice completes in approximately $22$~s because it reuses existing Core \acp{NF} (no new \ac{AMF}/\ac{SMF}/\ac{UPF} instantiation). \ac{URLLC} (${\approx}53$~s) and \ac{mMTC} (${\approx}42$~s) require additional control-plane or stateful data-store preparation. The non-3GPP slice (${\approx}50$~s) adds \ac{N3IWF} and IPsec setup overhead. These timing differences follow directly from the per-scenario \ac{NF} placement, reuse, and isolation choices in Table~\ref{tab:scenario-mapping}, confirming that the design workflow (S-\ac{NSSAI} assignment, \ac{NF} selection/sizing, resource allocation) is efficient and scalable.

\begin{figure}[h!]
    \centering
    \includegraphics[width=\linewidth]{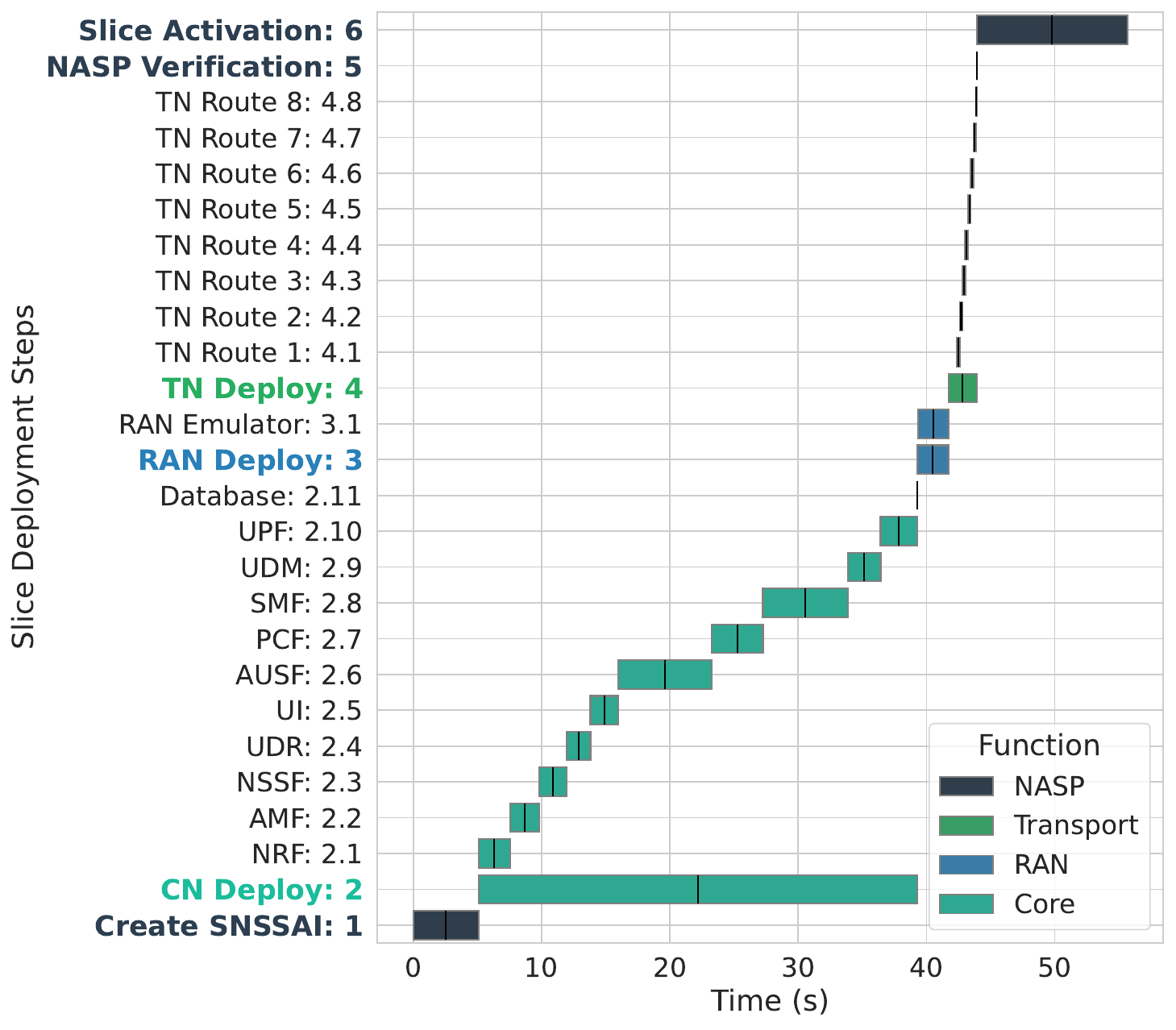}
    \caption{Per-step deployment time across network domains (\ac{CN}, \ac{RAN}, \ac{TN}).}
    \label{fig:slice_deploy_steps}
\end{figure}

The slice-instantiation workflow is decomposed into $N{=}26$ ordered steps and substeps (Figure~\ref{fig:slice_deploy_steps}). Step~1 creates the S-\ac{NSSAI} identifier and triggers the initial \ac{CN} deployment batch. Step~2 completes the \ac{CN} control-plane instantiation. Step~3 covers the \ac{RAN} domain, where only an emulated \ac{RAN} is considered via my5G-RANTester. Step~4 schedules the eight full-duplex \ac{TN} routes (four SDN hops $\times$ two directions) that connect the \ac{RAN} and \ac{CN} across the emulated WAN. Steps~5--6 execute \ac{NASP} verification and issue the slice-activation handshake. We define per-step efficiency as the ratio between the number of provisioning operations executed within a step (e.g., Helm-chart releases, flow-rule insertions, database transactions) and the elapsed time of that step, expressed in actions~$\text{s}^{-1}$. A larger value therefore indicates that the orchestrator completes more work per unit time, rather than merely the step being short. The step-wise breakdown suggests that reducing the number of \acp{NF} redeployed, particularly in Step~2, can lower the overall provisioning time. Step~2 corresponds to the \ac{CN} phase in which \ac{CN}~\ac{NSSMF} instantiates or reconfigures the control plane. Because this stage involves (i) pulling container images, (ii) applying Kubernetes manifests, and (iii) waiting for readiness probes across dependent microservices, it accounts for roughly 40\,\% of the total deployment time.

Figure~\ref{fig:slice_deploy_steps} breaks down the contribution of each deployment step per domain. The \ac{CN} domain exhibits the longest deployment times due to \ac{NF}-to-\ac{NF} internal communication and the overhead of Helm chart instantiation. The \ac{RAN} domain exhibits intermediate delays, driven by the my5G-RANTester emulation components that initialize the logical \textit{gNB} pods and execute streamlined configuration scripts. By contrast, the \ac{TN} domain exhibits more agile behavior due to its direct interaction with its dedicated controller. This phase involves configuring the eight full-duplex \ac{TN} routes (4 hops × 2 directions), i.e., all flow rules required by the emulated \ac{TN} topology. These observations reinforce that streamlined control domains, such as \ac{TN}, benefit from specialized controllers that minimize latency and improve overall efficiency of deployment.

To further contextualize the provisioning results, we relate NASP's \ac{E2E} slice deployment time (22--53~s, Fig.~9) and stepwise orchestration breakdown (Fig.~10) to timings from prior studies with different scopes. The sub-200~ms instantiation reported by \cite{Scotece2023-5gkube} refers to localized cloud-native \ac{CNF}/core startup. This evaluation serves only as a contextual lower-bound reference, not a direct \ac{E2E} comparator. In contrast, \cite{Dalgitsis2024-kp} reports strategy-dependent slice deployment times of 27.0, 19.4, 21.9, \(\approx 4\), and \(\approx 4\)~s in a federated multi-AD cloud-native 5G setup (post-federation deployment under different NF sharing/isolation strategies), which provides a partial comparator for NASP's 22--53~s \ac{E2E} provisioning. Quantitatively, these results place both studies in the seconds-to-tens-of-seconds range when orchestration scope extends beyond isolated function startup, and show that deployment time varies materially with workflow structure and sharing/isolation choices. We use the 33--38~ms (London) and 45~ms (Ireland) cloud-region latency values reported by \cite{Dalgitsis2024-kp} only as environmental context for edge-cloud experiments, not as direct slice-\ac{KPI} outcomes, to avoid conflating network-path conditions with NASP's service-level latency measurements.

\subsection{Lifecycle Management and Dynamic Reconfiguration}
\begin{figure}[h!]
    \centering
    \includegraphics[width=\linewidth]{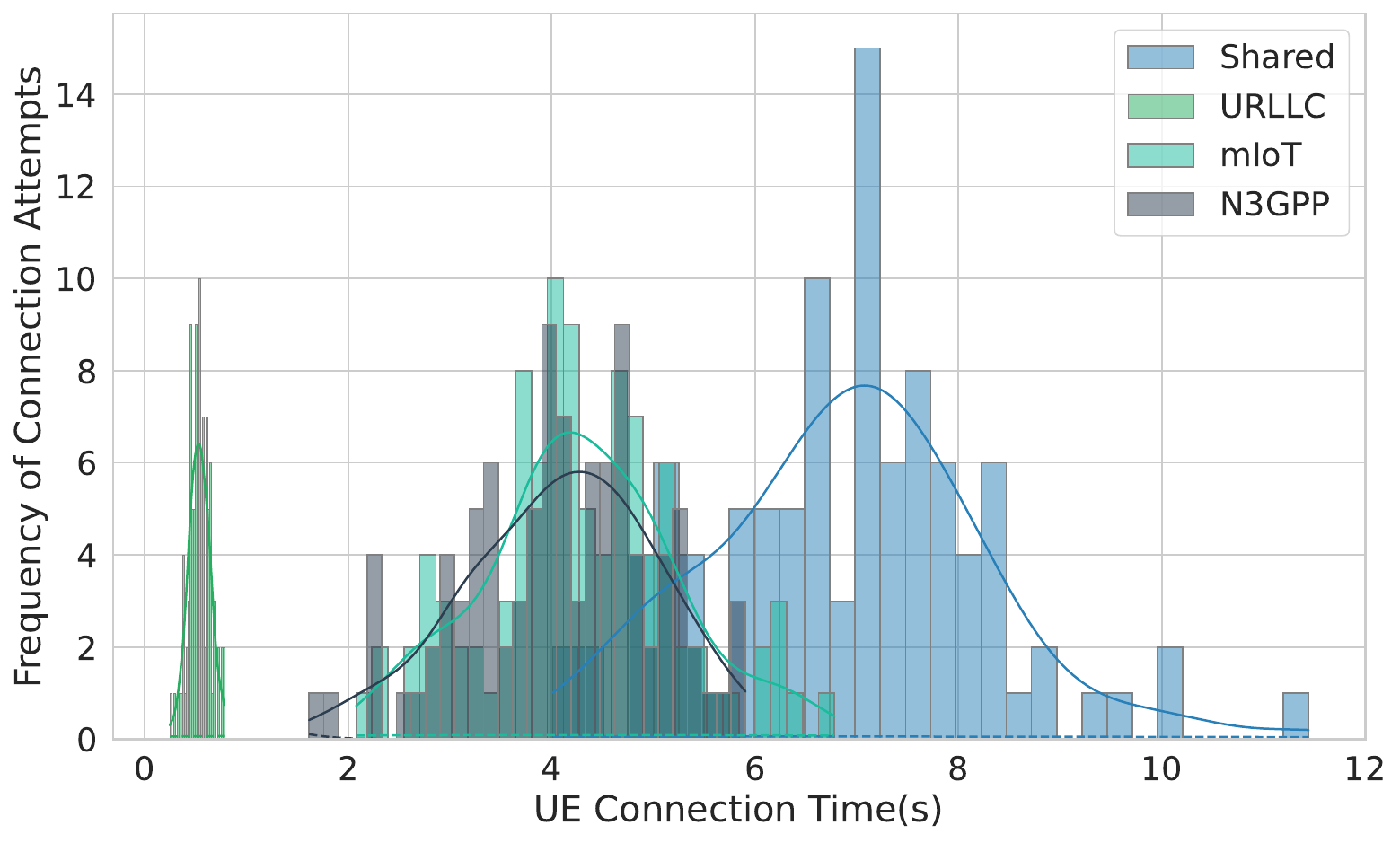}
    \caption{UE connection time across four scenarios (architectural mapping in Figure~\ref{fig:architecture} and TN-route types in Figure~\ref{fig:transport-topology}).}
    \label{fig:ue-conn-time}
\end{figure}

The slice-lifecycle evaluation investigates \ac{UE} attachment latency, dynamic slice reconfiguration, and resource utilization. Figure~\ref{fig:ue-conn-time} shows the distribution of \ac{UE} connection latencies in the four scenarios. These latency profiles are the result of the scenario-specific \ac{NF} placement (Table~\ref{tab:scenario-mapping}), routing choices (Figure~\ref{fig:transport-topology}), and orchestrator sequencing. The \ac{URLLC} scenario registers the lowest latency, due to the proximity of critical control functions and an exclusive \ac{TN} route. The \textit{mMTC} (mIoT) slice exhibits a moderate median with a pronounced tail caused by random-access congestion among many battery-constrained devices. By contrast, the Shared scenario incurs higher latency due to resource sharing among \acp{UE}.

The arrangement observed in Figure~\ref{fig:ue-conn-time} occurs from four specific NASP design decisions:  
(i) URLLC keeps \ac{AMF}/\ac{SMF} replicas at the internal-edge site and installs a short, dedicated \ac{VLAN} path (Figure~\ref{fig:transport-topology}), holding the control-plane \ac{RTT} below 3~ms. (ii) mIoT uses the same edge placement, but random-access congestion in the RAN dominates join latency, creating the long tail. (iii) Shared multiplexes \ac{CN} functions and \ac{TN} routes across slices, so queueing inside common \ac{CN} pods inflates the median attach time. (iv) Non-3GPP adds an extra N3IWF gateway, along with an IPsec tunnel setup (step 4 in Figure~\ref{fig:slice_deploy_steps}), which inserts roughly 1–1.5~s before \ac{AMF} messages reach the \ac{CN}. Moreover, the \textit{non-3GPP} slice experiences additional attachment delay from IPsec tunnel establishment and interworking procedures required for offloading traffic from untrusted access. Values beyond \(\pm3\sigma\) were discarded as outliers to strengthen statistical significance and, hence, the reliability of the reported trends.

\begin{figure}[!h]
    \centering
    \includegraphics[width=\linewidth]{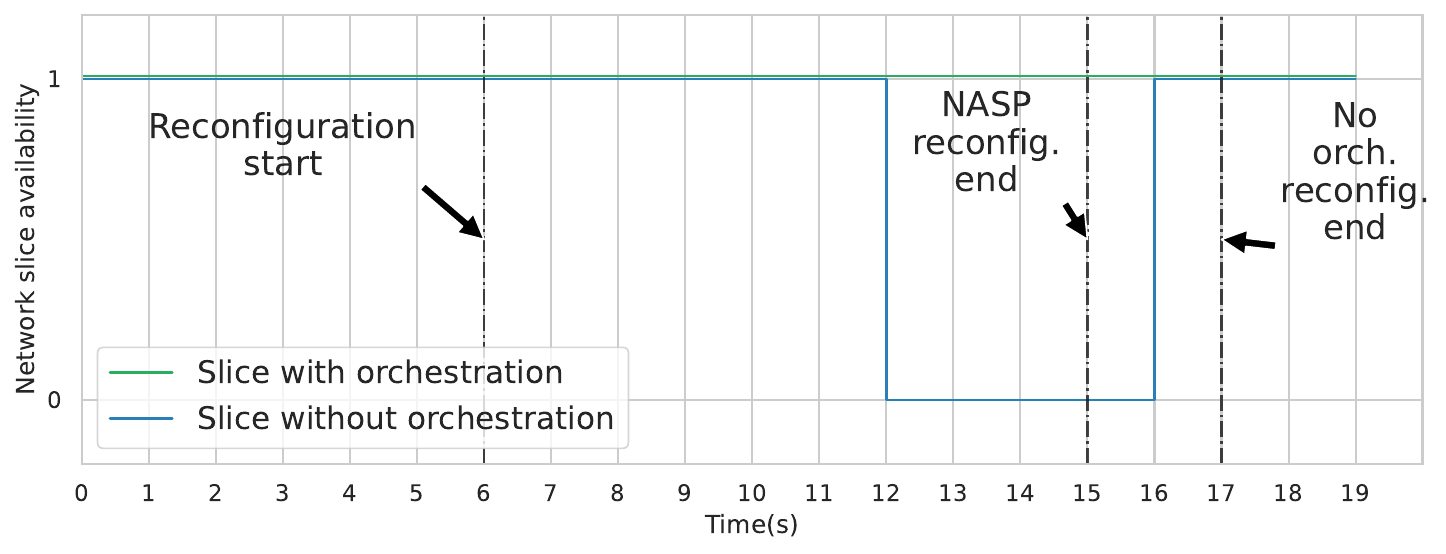}
    \caption{Slice availability during runtime \ac{AMF} replace-and-redeploy procedure: the controller deletes the old \ac{AMF} pod and launches a new one with an upgraded resource profile. The five vertical arrows indicate, in order, the start of Step~2.2, the outage interval while Step~2.2 executes, the end of Step~2.2, the completion of Steps~2.3–2.10, and the first successful \ac{UE} registration in Step~6 (see Figure~\ref{fig:slice_deploy_steps}).}
    \label{fig:disponibilidade}
\end{figure}

Dynamic reconfiguration was characterized by monitoring slice availability (a binary metric, where $0$ represents service outage and $1$ represents full operability) and UE registration latency during runtime updates of the \ac{CN} and its supporting virtual infrastructure.
The specific reorganization analyzed here transparently migrates the slice from an initial, low‑capacity control‑plane deployment to a higher‑capacity configuration that meets a sudden surge in \ac{QoS} and throughput requirements while keeping ongoing sessions untouched. Figure~\ref{fig:disponibilidade} shows that \ac{NASP} orchestration maintains service continuity even during the 9~s reconfiguration window delimited by the vertical dashed lines.

\begin{figure}[h!]
    \centering
    \includegraphics[width=\linewidth]{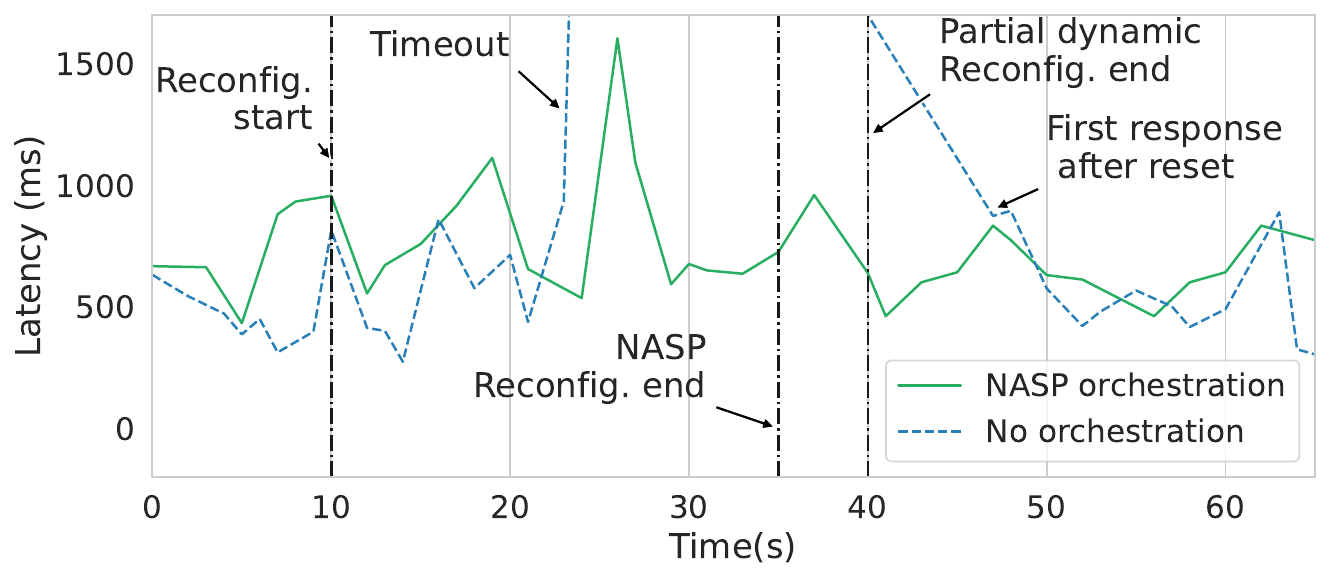}
    \caption{Registration latency during the reconfiguration process.}
    \label{fig:latency}
\end{figure}

We also analyze \ac{UE}-registration latency before, during, and after slice reconfiguration. 
Figure~\ref{fig:latency} shows the latency of the reconfiguration process with a 500~ms sampling period under \ac{NASP} orchestration. 
In this run, the latency peaks at 1500~ms and quickly returns to the pre-reconfiguration baseline of approximately 600~ms. The peak reflects limited computing resources and communication across multiple networks and subnetworks in the evaluation scenario.
The arrow labeled ``Reconfig.\ start'' (10~s) indicates when the controller begins destroying and redeploying the \ac{AMF} pods. The subsequent arrow, ``Timeout'' (23~s), marks when the \ac{UE} registration procedure stops receiving responses because no \ac{AMF} instance is reachable. Service availability is restored at ``NASP reconfig.\ end'' (30~s) when the new \ac{AMF} replica becomes ready. At 40~s, ``Partial-dynamic reconfig.\ end'' marks the completion of the partial dynamic orchestration, and ``First response after reset'' (47~s) highlights the first successful \ac{UE} registration after the process. Destroying and redeploying the \ac{AMF} pod is necessary because the new configuration, such as the higher replica count and tightened CPU quotas, modifies immutable Kubernetes fields. Therefore, Helm performs a recreate rather than an in-place update so that the fresh instance boots with the correct resource profile and slice identifiers. Since UE and session contexts reside in the external \ac{UDR}, the \ac{AMF} is stateless, and the brief $\approx9$\,s outage in Figure~\ref{fig:disponibilidade} does not affect the data plane. Moreover, Figure~\ref{fig:latency} shows the ``No orchestration'' latency. After reconfiguration starts at 10~s, latency rises sharply. No successful registration responses are observed between 23~s and 47~s, the exact interval highlighted by the ``Timeout'' arrow, because all attempts expire while the control plane is rebuilding.

We contextualize the runtime behavior observed in NASP against representative cloud-native slicing and core-network studies, while preserving the original metric semantics of those works. \cite{zhao2025} reports \(64.2\%\) total operating-cost reduction and \(45.5\%\) normalized-performance improvement, as well as converged orchestration outcomes of \(4.3/1.79\) (cost / normalized performance) versus \(12/1.23\) for Atlas, and dynamic adaptation after slice departure with \(25.6\%\) cost reduction and \(12.8\%\) performance improvement in 1 slot. These results provide a partial/contextual reference for NASP's runtime adaptation, cost-efficiency, and service-quality-oriented evaluation dimensions. \cite{zhao2025} also reports up to \(53.9\%\) degradation (normalized performance) for comparison systems beyond 3 slices (up to 5 slices), which we use as a partial scalability reference for multi-slice concurrency effects, although the slice abstractions and resource models differ from NASP~\cite{zhao2025}.

Likewise, CoreKube (\cite{Larrea2023}) reports control-plane scalability thresholds where NextEPC saturates beyond \(\approx 12\) requests/s, Open5GS saturates after \(\approx 120\) requests/s, and CoreKube scales beyond; a burst-handling result of 560~msgs/s from 250~\acp{UE}; resilience experiments with a 200~s critical-error interval (about \(\sim 20\times\) more frequent than the referenced baseline); and \ac{AMF} CPU imbalance of \(91.4/1.4/1.4\%\) across three \ac{AMF} pods in the K8s-Open5GS case. These results provide contextual evidence for \ac{CN}/control-plane scalability and robustness, and quantitatively demonstrate that replication alone does not ensure balanced scaling without adequate traffic distribution and orchestration. Finally, we explicitly note that the disruption interval discussed in the comparison setup corresponds to the baseline environment without NASP and is not attributed to NASP-managed \ac{AMF} redeployment. This part of the discussion presents a contextual comparison of reconfiguration/failure models and automation behavior rather than a direct outage benchmark.

\begin{figure}[h!]
    \centering
    \includegraphics[width=\linewidth]{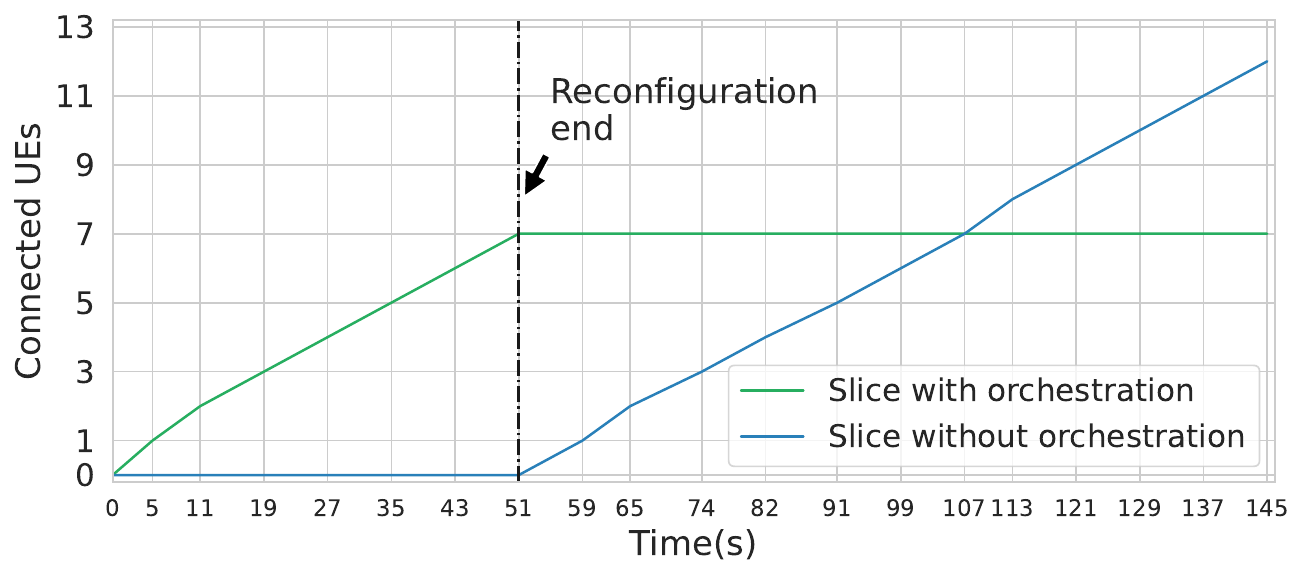}
    \caption{Slice utilization before and after dynamic reconfiguration (admission control event at dashed line).}
    \label{fig:uso}
\end{figure}

We assessed the adaptability of slice management by examining slice‑transition events triggered when the number of concurrent \ac{UE} connections exceeded a predefined threshold. Figure~\ref{fig:uso} shows that \ac{NASP} triggers reconfiguration to maintain \ac{QoS} without interruption. Until $t \simeq 51$\,s, every new UE is admitted to ``Slice with orchestration'', and its utilization grows linearly until it reaches the configured cap of seven simultaneous connections. At this instant, the \ac{NSSMF} controller finalizes the reconfiguration (vertical dashed line) and redirects subsequent registration requests to ``Slice without orchestration''. From that moment on, the green curve saturates while the blue curve rises, evidencing the controller’s ability to preserve the \ac{SLA} of the saturated slice while still accommodating additional traffic. The different slopes of the curves follow directly from the admission‑control policy rather than any performance disparity between the slices. This dynamic switch indicates that the \ac{NASP} architecture can support runtime scalability, ensuring that high-demand scenarios do not compromise overall network performance.  

\subsection{Resource Utilization and Cost Analysis}
\begin{figure}[h!]
    \centering
    \includegraphics[width=\linewidth]{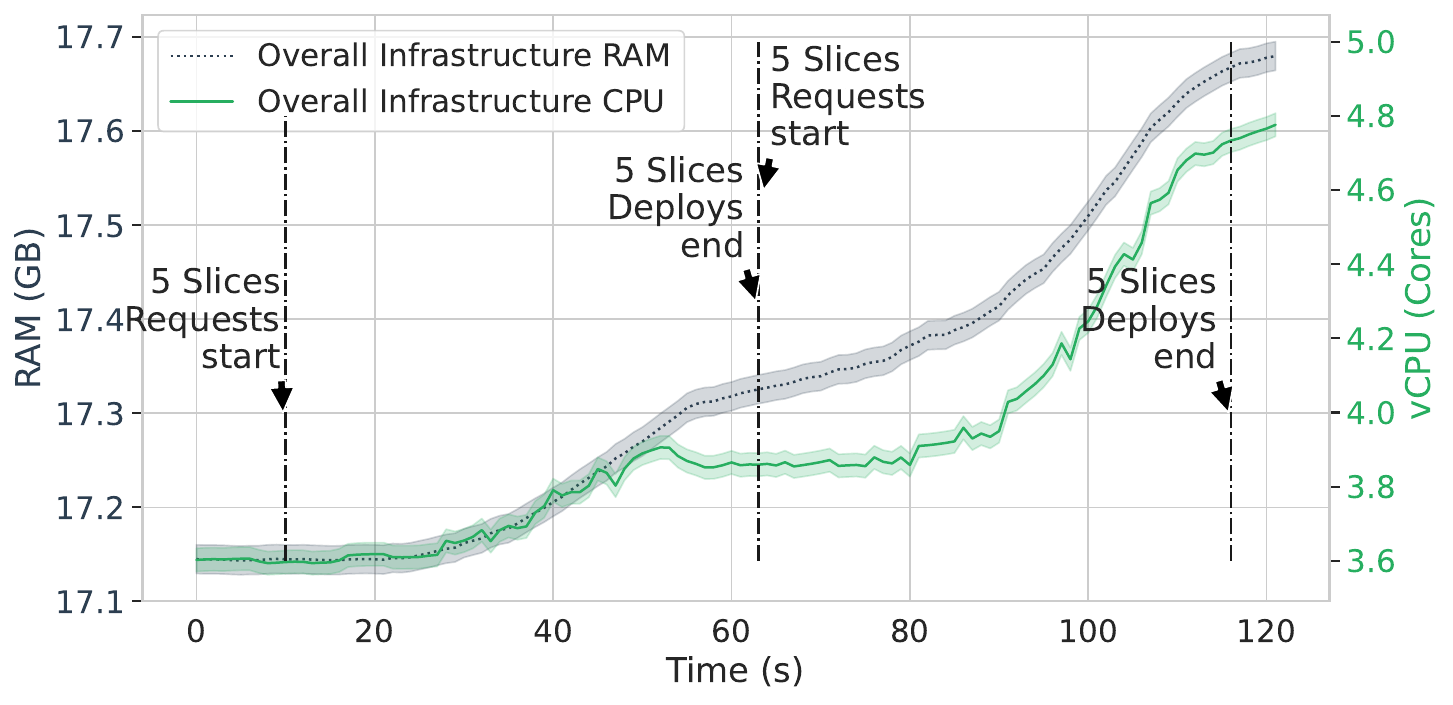}
    \caption{\ac{vCPU} and \ac{RAM} utilization over the \ac{NSI} lifecycle.}
    \label{fig:slice-usage}
\end{figure}

We also monitored resource utilization continuously during slice deployment. Figure~\ref{fig:slice-usage} presents combined \ac{vCPU} and \ac{RAM} usage across the testbed, with averages of approximately 1.2~\ac{vCPU} and 600~MB \ac{RAM} over the \ac{NSI} lifecycle. Monitoring these metrics highlights the efficiency of resource allocation and the small fluctuations resulting from automation-induced bursts, which are crucial for fine-tuning cost-management strategies in different deployment environments. The three vertical dash‑dotted arrows superimposed on Figure~\ref{fig:slice-usage} mark resource demand: (i) the left-most arrow identifies when the batch of five slice requests is submitted, (ii) the central arrow marks the transition from request processing to slice instantiation, and (iii) the right-most arrow denotes when all five slices are fully deployed, at which point the infrastructure peaks at about 4.8\,\ac{vCPU} and 17.6\,GB of RAM. These markers help relate each lifecycle stage to its corresponding CPU and memory consumption.

Finally, we estimate the evaluation cost by relating the consumed computational resources to their operating costs in each environment (Edge, Metropolitan, Central (Cloud)). Unit prices were taken from public, on‑demand Amazon Web Services tariffs and summarized in Table~\ref{tab:instance-price}. For each environment, we selected the closest general‑purpose flavor available in that tier. Monthly prices were normalized by advertised resources so that the regression coefficients express the marginal cost, in $\$\,\text{month}^{-1}$, of one \ac{vCPU} and one GiB of \ac{RAM}.

\begin{table}[!h]
\caption{Description of Edge, Metropolitan, and Central tier instances~\citep{amazonStartedWith}.}
\label{tab:instance-price}
\centering
\resizebox{\linewidth}{!}{%
\begin{tabular}{l|c|c|c|c|c}
\hline
\rowcolor[HTML]{EFEFEF}
\textbf{Type} & \textbf{Size} & \textbf{vCPU} &
  \makecell{\textbf{RAM}\\\textbf{(GB)}} &
  \makecell{\textbf{Storage}\\\textbf{(GB)}} &
  \textbf{Price/Month} \\ \hline
Edge           & medium  & 2 & 4  & 200 & \$70,88  \\ \hline
Edge           & xlarge  & 4 & 16 & 200 & \$193,52 \\ \hline
Edge           & 2xlarge & 8 & 64 & 200 & \$526,40 \\ \hline
Metropolitan   & medium  & 2 & 4  & 200 & \$67,96  \\ \hline
Metropolitan   & xlarge  & 4 & 16 & 200 & \$117,60 \\ \hline
Metropolitan   & 2xlarge & 8 & 64 & 200 & \$181,84 \\ \hline
Central        & medium  & 2 & 4  & 200 & \$46,37  \\ \hline
Central        & xlarge  & 4 & 16 & 200 & \$76,74  \\ \hline
Central        & 2xlarge & 8 & 64 & 200 & \$137,47 \\ \hline
\end{tabular}%
}
\end{table}

\begin{figure}[h!]
    \centering
    \includegraphics[width=\linewidth]{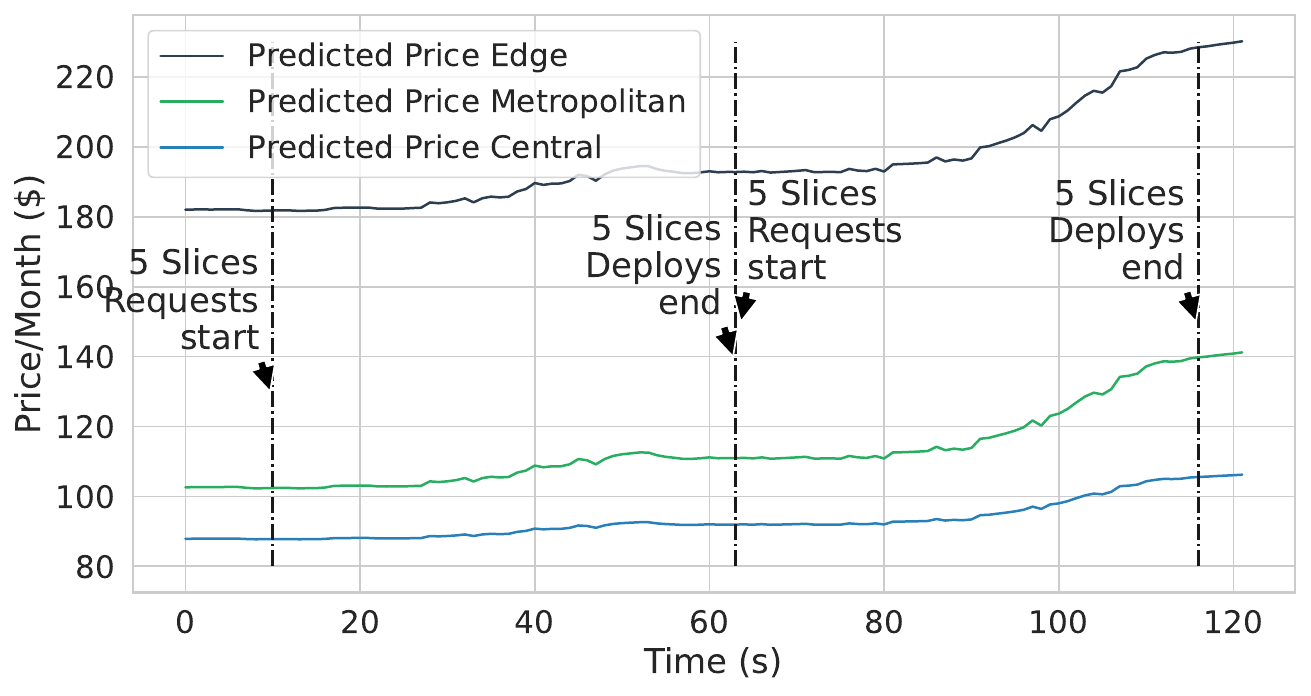}
    \caption{Predicted monthly cost as a function of \ac{vCPU} and \ac{RAM} utilization.}
    \label{fig:slice-price}
\end{figure}

We fit a linear regression to the values in Table~\ref{tab:instance-price}, yielding one equation per environment: $\text{Edge}=39.42\,\text{CPU}+3.65\,\text{RAM}-22.56$, $\text{Metropolitan}=33.58\,\text{CPU}-1.46\,\text{RAM}+6.63$, and $\text{Central}=15.19\,\text{CPU}+1\,\text{RAM}+15.99$. We then used the consumption data from the previous evaluation to predict the price for each environment. As shown in Figure~\ref{fig:slice-price}, the dash–dotted vertical lines mark (i) the submission of the first batch of five slice requests at $t = 10$~s, (ii) the completion of that batch’s deployment at $t \approx 65$~s, (iii) the submission of a second batch of five slice requests at $t = 60$~s, and (iv) the instant when all ten slices are fully operative at $t \approx 115$~s. Each marker corresponds to an inflection point in the Edge, Metropolitan, and Central cost curves, illustrating how incremental \ac{vCPU} and \ac{RAM} allocations immediately increase the predicted monthly costs.

These experimental results validate the capability of the \ac{NASP} architecture to efficiently support network‑slice design, deployment, and dynamic lifecycle management. Beyond demonstrating flexibility across four \ac{5G} scenarios, the evaluation indicates a balanced trade‑off between deployment speed, run‑time performance, and operational cost. Note that our current evaluation environment uses the my5G-RANTester for \ac{RAN} emulation, with a focus on end-to-end orchestration workflows. This is because \ac{NASP} operates at the orchestration layer, managing the lifecycle at the network-function level rather than at the individual-flow or physical-link level. 
Mobility and handover procedures are executed within the \ac{RAN} execution layer, whereas \ac{NASP} interfaces with \ac{RAN} components at the management and orchestration layer. While mobility and large-scale attachment effects influence runtime service performance and \ac{QoS} metrics, they do not affect the architectural validity of the orchestration mechanisms evaluated in this work. In this context, \ac{NASP} provides a viable platform for next‑generation network slicing.

The experiments consider up to 250 concurrent \acp{UE} per slice, and our experimental topology cannot fully capture topology-driven effects such as distributed domain coordination, transport-domain expansion, and end-to-end operational delays. This is because the evaluation focuses on the scalability and efficiency of the control plane and the orchestration lifecycle (e.g., slice instantiation, configuration, and end-to-end lifecycle management) in a controlled proof-of-concept environment, rather than on stressing data-plane capacity typical of large-scale \ac{mMTC} deployments. In our framework, the orchestration overhead is primarily determined by the number of managed slices, instantiated network functions, and lifecycle events, rather than directly by the number of attached devices or the raw number of switches or physical links, although the scale and heterogeneity of the topology may still influence the practical orchestration latency. Therefore, the selected experimental scale enables us to capture the relevant behavior of the orchestration mechanisms as slice management complexity increases. While \ac{mMTC} scenarios typically involve much larger device populations, the goal of this study is to evaluate the orchestration framework's ability to efficiently deploy, manage, and coordinate multiple network slices. Future work will extend the evaluation to larger device populations and heterogeneous, geographically distributed topologies, including real-world \ac{RAN} environments and mobility scenarios, to further assess the system under large-scale \ac{mMTC} conditions.
\section{Conclusion}\label{sec:conclusion}

The evolution of \ac{5G} mobile networks is catalyzing digital transformation across many sectors, with standardization bodies such as \ac{ETSI} and \ac{3GPP}, driving major architectural and technological advances. In this context, \ac{NSaaS} is increasingly adopted as a foundation for flexible, tenant-oriented connectivity; however, its \ac{E2E} realization still lacks a holistic standard. This work addresses this gap by proposing the \ac{NASP} architecture, which maps business-level \ac{GSMA} templates to technical \ac{NSI} definitions and establishes connectivity across the \ac{RAN}, \ac{TN}, and \ac{CN} domains. \ac{NASP} is designed to interact with domain controllers, such as \ac{K8s} and \ac{ONOS}, and with standard Linux tools, such as \texttt{iptables}, namespaces, and \ac{CGROUPS}, to enforce configuration changes, for example namespace routing and \ac{VLAN} creation.

Evaluation across four scenarios demonstrates the platform’s capability to translate \ac{SLA} specifications into \ac{E2E} slice configurations, automate slice instantiation and management, and balance adaptability with resource consumption. Three quantitative findings are notable: (i) approximately 66\% of the instantiation time is incurred in the \ac{CN} domain, (ii) the \ac{URLLC} slice achieves a 93\% reduction in data-session establishment time compared with the Shared slice, and (iii) there is a 112\% difference in monthly cost between Edge and Centralized deployments under identical performance targets.

Beyond the architectural contributions, the seamless integration of standardized \ac{GSMA} templates, the unification of definitions across major mobile network entities to realize an \ac{E2E} architecture, and the capability to instantiate network slices for \ac{non3GPP} applications open several avenues for future work. Planned efforts include: (a) integrating real-world physical \ac{RAN} deployments to assess radio-level effects and mobility procedures that were abstracted in this study, (b) scaling the physical topology across heterogeneous domains to stress-test the data plane under massive machine-type communications (\ac{mMTC}) limitations, (c) embedding advanced \ac{AI}-driven analytics for predictive scaling and anomaly classification within the closed-loop assurance mechanisms, and (d) validating interoperability with commercial \ac{MANO} stacks, such as \ac{ONAP} and \ac{OSM}, while aligning interfaces with emerging \ac{O-RAN} specifications in multi-operator environments.

\section*{Acknowledgment}

This work was partially supported by CNPq Grants Nos.\ 405111/2021-5 and 130555/2019-3, and by CAPES, Finance Code 001, Brazil; additional support was provided by RNP and MCTIC under Grant No.\ 01245.010604/2020-14 as part of the 6G Brasil and OpenRAN@Brasil projects, and by MCTIC/CGI.br/FAPESP through Project SAMURAI (Grant No.\ 2020/05127-2) and Project PORVIR-5G (Grant No.\ 2020/05182-3); it has also been partially funded by the project XGM-AFCCT-2024-5-1-1 supported by xGMobile - EMBRAPII - Inatel Competence Center on 5G and 6G Networks, with financial resources from the PPI IoT/Manufatura 4.0/the MCTI grant number 052/2023, signed with EMBRAPII.
%% Bibliography

\bibliographystyle{elsarticle-harv}
\bibliography{ref.bib}

\end{document}